\documentclass[journal,11pt,draftclsnofoot,onecolumn]{IEEEtran} 

\usepackage[caption=false,font=normalsize,labelfont=rm,textfont=rm]{subfig}
\usepackage{algorithm}
\usepackage{algorithmic}
\usepackage{bm}
\usepackage{booktabs}
\usepackage{amsthm}
\usepackage{amsmath}
\usepackage{stfloats}
\usepackage{amsfonts,amssymb}   
\usepackage[utf8]{inputenc}
\usepackage{graphicx}
\usepackage{epstopdf}
\usepackage{cite}
\usepackage{multirow}
\usepackage[T1]{fontenc}
\usepackage{caption}
\usepackage{float}            
\usepackage{subfloat}
\usepackage{setspace}
\usepackage{color}
\usepackage{easyReview}
\allowdisplaybreaks

\begin{document}

\title{Transmit Beampattern Synthesis for Active RIS-Aided MIMO Radar via Waveform and Beamforming Optimization} 

\author{Shengyao Chen, Minghui He, Longyao Ran, Hongtao Li, Feng Xi, Sirui Tian, Zhong Liu

\thanks{
		
The authors are with the School of Electronic and Optical Engineering, Nanjing University of Science and Technology, Nanjing 210094, China	(e-mail: chenshengyao@njust.edu.cn, rly@njust.edu.cn, xifeng@njust.edu.cn, liht@njust.edu.cn, tiansirui@njust.edu.cn, eezliu@njust.edu.cn). 
}
}

\maketitle

\begin{abstract}
In conventional colocated multiple-input multiple-output (MIMO) radars, practical waveform constraints including peak-to-average power ratio, constant or bounded modulus lead to a significant performance reduction of transmit beampattern, especially when the element number is limited.
This paper adopts an active reconfigurable intelligent surface (ARIS) to assist the  transmit array and discusses the corresponding beampattern synthesis.	
We aim to minimize the integrated sidelobe-to-mainlobe ratio (ISMR) of beampattern by the codesign of waveform and ARIS reflection coefficients.
The resultant problem is nonconvex constrained fractional programming whose objective function and plenty of constraints are variable-coupled.
We first convert the fractional objective function into an integral form via Dinkelbach transform, and then alternately optimize the waveform and ARIS reflection coefficients.
Three types of waveforms are unifiedly optimized by a consensus alternating direction method of multipliers (CADMM)-based algorithm wherein the global optimal solutions of all subproblems are obtained, while the ARIS reflection coefficients are updated by a concave-convex procedure (CCCP)-based algorithm. 
The convergence is also analyzed based on the properties of CADMM and CCCP.
Numerical results show that ARIS-aided MIMO radars have superior performance than conventional ones due to significant reduction of sidelobe energy.

\end{abstract}


\IEEEpeerreviewmaketitle


\section{Introduction}

\IEEEPARstart{O}{wing} to plentiful degrees-of-freedom (DoFs) generated by waveform diversity, colocated multiple-input multiple-output (MIMO) radars have great flexibility in signal processing and thus harvest unique performance advantages.
As a vital technique, MIMO waveform design has received widespread attention during the past decades \cite{Stoica2007OnProbing,Stoica2008Waveform,Aubry2016MIMO,Zhao2017Aunified,Cheng2019Co-Design,Stoica2012Optimization,Leshem2007Information,Chen2013Adaptive,Cui2014MIMO,Ahmed2014MIMO,Cui2017Space,Kay2009Waveform}.
Based on different radar performance requirements, it can be classified into the following four categories:
1) Enhance target detection performance \cite{Stoica2007OnProbing,Stoica2008Waveform,Aubry2016MIMO} or target discrimination \cite{Zhao2017Aunified} via transmit beampattern synthesis;
2) Improve parameter estimation accuracy by reducing Cramer-Rao bound (CRB) \cite{Cheng2019Co-Design} or mean square error (MSE) \cite{Stoica2012Optimization};
3) Boost target recognition performance by maximizing mutual information (MI) \cite{Leshem2007Information,Chen2013Adaptive};
4) Improve target detection performance by maximizing output signal-to-interference-plus-noise ratio (SINR) \cite{Cui2014MIMO,Ahmed2014MIMO,Cui2017Space} or relative entropy\cite{Kay2009Waveform} in combination with receive filter design.
This paper focuses on the transmit beampattern synthesis of colocated MIMO radars.


In the initial stage, the beampattern synthesis is often conducted by two-step methods, where the waveform covariance matrix synthesis and MIMO waveform design are sequentially accomplished \cite{Stoica2007OnProbing,Stoica2008Waveform,Aubry2016MIMO}.
For example, Stoica et al. used semidefinite quadratic programming (SQP) to optimize the covariance matrix based on beampattern matching or sidelobe level (SLL) minimization criteria \cite{Stoica2007OnProbing}, and then offered an efficient cyclic algorithm (CA) to design the MIMO waveforms obeying peak-to-average power ratio (PAR) or constant modulus (CM) constraints \cite{Stoica2008Waveform}.
Aubry et al. generated the robust covariance matrix by minimizing the peak sidelobe level (PSL) or integral sidelobe level (ISL) in virtue of semidefinite programming (SDP) \cite{Aubry2016MIMO}.
However, these two-step methods inevitably result in a considerable performance loss due to indirect design.

To remedy this issue, several direct methods are developed with the aid of advanced nonconvex optimization techniques \cite{Wang2012On,Cheng2017Constant,Fan2018Constant,Cheng2018Communication,Zhao2018MIMO,Zhou2019Unified,Fan2019MIMO,Imani2021ACoordinate,Raei2022MIMO}.
Wang et al. first minimized the quartic beampattern matching error under CM constraints by using L-BFGS \cite{Wang2012On}. 
Thereafter, Cheng et al. adopted the alternating direction method of multipliers (ADMM) to handle the similar problems \cite{Cheng2017Constant}.
Due to its universality, the ADMM is then respectively applied to minimize the PSL \cite{Fan2018Constant} and integrated sidelobe-to-mainlobe ratio (ISMR) \cite{Cheng2018Communication}.
On the other side, majorization-minimization (MM) is another popular method for solving nonconvex problems thanks to its fast convergence speed and low complexity \cite{Sun2017MM}. 
In \cite{Zhao2018MIMO} and \cite{Zhou2019Unified}, the authors gave a unified MM-based framework to minimize the beampattern matching errors under several types of waveform constraints.
A weighted $l_p$-norm-based metric is then employed to yield quasi-equiripple beampatterns in \cite{Fan2019MIMO}.
Furthermore, coordinate descent (CD) \cite{Imani2021ACoordinate} and block successive upper-bound minimization (BSUM) \cite{Raei2022MIMO}-based approaches were established to deal with discrete phase waveform constraints.


It is notable that conventional colocated MIMO radars often require a lot of transmit channels to provide sufficient spatial DoFs for achieving good beampattern performance.
In some military or civilian applications, such as unmanned aerial vehicle-mounted and automotive radars, the size and DoFs of MIMO arrays are significantly limited by physical constraints such as cost and volume, making it difficult to achieve high beampattern performance.
In view of this, equipping a reconfigurable intelligent surface (RIS) with the transmit array is a viable scheme to boost the MIMO beampattern performance since an RIS can introduce plenty of additional DoFs without altering the MIMO system structure.
As we know, an RIS is a novel antenna array consisting of multiple metasurface reflecting elements, and it has the capability of ingeniously manipulating wireless environments by tuning the phase shifts of reflected signals \cite{Renzo2020Smart}.
Owing to its merits including low cost, low power consumption and flexible configuration, RISs have become one of the fundamental technologies for next-generation wireless communications. 
Now, it is extensively used in enhancing wireless communication performance, expanding coverage range, ensuring physical layer security, simultaneous wireless information and power transfer (SWIPT), and other areas \cite{Liu2021Reconfigurable}.



Inspired by its immense advantages in wireless communications, RISs have recently been utilized in different MIMO radars to enhance their performance as well \cite{Cisija2021RIS,Meng2022Intelligent,Esmaeilbeig2023Joint,Chen2023Robust,Grossi2023Beampattern,Lu2021Target,Lu2021Intelligent,Buzzi2022MIMO,Liu2023Joint}.
In scenarios where line-of-sight (LoS) paths are blocked, an RIS enables MIMO probing signals to propagate through non-line-of-sight (NLoS) paths,
not only achieving multi-target localization \cite{Cisija2021RIS} and sensing \cite{Meng2022Intelligent} but also improving parameter estimation accuracy \cite{Esmaeilbeig2023Joint,Chen2023Robust}.
It is further applied to transmit beampattern synthesis of wideband MIMO for reducing the number of transmit channels \cite{Grossi2023Beampattern}.
On the other side, when LoS paths are unobstructed, RISs still have the capability of boosting MIMO radar performance \cite{Lu2021Target,Lu2021Intelligent,Buzzi2022MIMO,Liu2023Joint}.
For instance, Lu et al. deployed an RIS close to distributed or colocated MIMO radar receivers to enhance the received signal-to-noise ratio (SNR) \cite{Lu2021Target, Lu2021Intelligent}.
Buzzi et al. equipped the transmitter and receiver of bistatic MIMO radar respectively with a closely-spaced RIS to improve the target detection performance \cite{Buzzi2022MIMO}.
Moreover, Liu et al. employed multiple RISs to assist a colocated MIMO radar for multi-target detection with each RIS near a target and effectively suppressed interferences by the codesign of transmit waveform, RISs and receive filters~\cite{Liu2023Joint}.


It is worth pointing out that all used RISs are passive in the aforementioned literature \cite{Cisija2021RIS,Meng2022Intelligent,Esmaeilbeig2023Joint,Chen2023Robust,Grossi2023Beampattern,Lu2021Target,Lu2021Intelligent,Buzzi2022MIMO,Liu2023Joint}.
Due to the impact of double fading effect, the passive RIS (PRIS) only yields a limited integration gain from the NLoS path because it can only adjust the phase shifts of reflected signals but cannot change their amplitudes, even though offering a great number of DoFs.
To tackle this issue, several types of simple amplification circuits are equipped with the PRIS element, prompting the birth of active RIS (ARIS) \cite{Zhang2023Active,Long2021Active,Rao2023Active}.		
Compared to PRISs, ARISs have the capability of simultaneously adjusting the amplitude and phase shift of incident signals, hence producing not only more DoFs but also a larger integration gain.
Currently, the merits of ARIS have been extensively utilized in wireless communications \cite{Long2021Active}, secure transmission \cite{Dong2022Secure}, integrated sensing and communication (ISAC) \cite{Zhu2023Joint}, SWIPT \cite{Gao2023Beamforming}, phased-array radars \cite{Ran2023Beampattern} and MIMO radars \cite{Zhang2024Active}, respectively.
In terms of transmit beampattern synthesis, Ran et al. adopted an ARIS to enable the PSL reduction of phased-array radars and thus achieved significantly superior performance to the PRIS-enabled one \cite{Ran2023Beampattern}. 
By minimizing the beampattern matching error with user SINR constraints, Zhang et al. synthesized a desired reflective beam for NLoS target sensing in an ARIS-aided ISAC \cite{Zhang2023ActiveIRS}.
However, there is a lack of the study on beampattern synthesis of ARIS-aided MIMO radars.

In view of the performance advantages offered by ARISs in phased-array radars and ISACs, this paper considers deploying a closely-spaced ARIS to enable the beampattern synthesis of colocated MIMO radars.
To reduce the transmit energy in the sidelobe region and concentrate more energy in the mainlobe region simultaneously, we use the ISMR as a metric and jointly design the transmit waveform and ARIS reflection coefficients.
Different from traditional colocated MIMO radars, the extra NLoS signal yielded by the ARIS can be utilized to weaken the illumination energy in the sidelobe region while enhance the  energy in the mainlobe region.
Therefore, we can effectively reduce the ISMR from these two aspects.
To achieve high transmit efficiency, we investigate several typical waveform constraints respectively, including CM, PAR and bounded modulus (BM) constraints.
We also consider the practical constraints of ARIS yielded by hardware implementations.
The main contributions of this paper include:

\begin{itemize}
	\item \emph{Problem Formulation:}
	Based on the criterion of ISMR minimization, we formulate the beampattern synthesis of ARIS-aided MIMO radar as a fractional biquadratic function minimization problem under nonconvex constraints. 
	Concretely, the MIMO waveform suffers from CM, PAR or BM constraints, and the ARIS reflection coefficients are limited by the maximum amplification factor of each element and the maximum reflecting power.
	As the objective function and all constraints except the amplitudes of ARIS reflection coefficient are nonconvex, the proposed problem belongs to nonconvex constrained fractional programming (FP), which is distinct from existing MIMO beampattern synthesis problems.
	
	\item \emph{Algorithm Design:} 
    We first supersede the fractional objective function with an equivalent integral form by leveraging Dinkelbach transform. 
    Then we propose an effective codesign algorithm under the alternating minimization (AM) framework. 
    Concretely, three types of transmit waveforms are uniformly optimized by a customized consensus ADMM (CADMM)-based algorithm in which the global optimal solutions of all subproblems are obtainable, while the ARIS reflection coefficients are updated by a concave-convex procedure (CCCP)-based algorithm which contains only several simple convex subproblems.
    Since all subproblems can be handled tractably, the proposed algorithm is an effective one.
		
	\item \emph{Performance Evaluations:} 
	Extensive numerical experiments verify the effectiveness of the proposed algorithm and the superior beampattern performance generated by the ARIS.
	With three types of practical waveforms, the proposed algorithm always converges and the deployed ARIS significantly reduces the ISMR.
	Compared to conventional MIMO radars, the ARIS dramatically reduces the illumination energy in the sidelobe region, hence markedly decreasing SLLs.
	It also slightly increases the energy in the mainlobe region and yields a more uniform energy distribution.
	With the increase of ARIS element number and maximum amplification factor, the ISMR performance is further enhanced.

\end{itemize}

$\emph{Notations}$:
Vectors and matrices are denoted by boldface lowercase and uppercase letters, respectively.
The $(\cdot)^{T}$ , $(\cdot)^{H}$, $(\cdot)^{*}$, $\angle(\cdot)$ and $|\cdot|$ are the transpose, conjugate transpose, complex conjugate, phase and modulus operators, respectively.
The trace, rank and stacking vectorization of a matrix are written as $\mathrm{tr}(\cdot)$, $\mathrm{rank} (\cdot)$ and $\mathrm{vec} (\cdot)$.
$\mathrm{Re}\{\cdot\}$ denotes the real component of a complex number.
$\mathbb{E}\{\cdot\}$ denotes the mathematical expectation.
$\bm{I}_{N}$ stands for an $N\times N$ identity matrix.
$\bm{X} \succeq 0$ means that the matrix $\bm{X}$ is positive semidefinite.
For a matrix $\bm{X} \in \mathbb{C}^{L\times N}$, 
$\bm{x}(n)$ denotes the vector of the $n$-th column of $\bm{X}$, 
and $x_l(n)$ denotes the element of $\bm{X}$ at the $n$-th column and $l$-th row.
For a vector $\bm{x}$, $x_n$ stands for the $n$-th entry of $\bm{x}$.
$\mathrm{Diag}(\bm{x})$ is a diagonal matrix with $\bm{x}$ filling its principal diagonal and $\mathrm{diag}(\bm{X})$ is the vector consisting of all diagonal elements of $\bm{X}$.
$\otimes$ and $\odot$ denote Kronecker product and Hadamard product.

\begin{figure}[!t]
	\centering
	\includegraphics[scale=0.47]{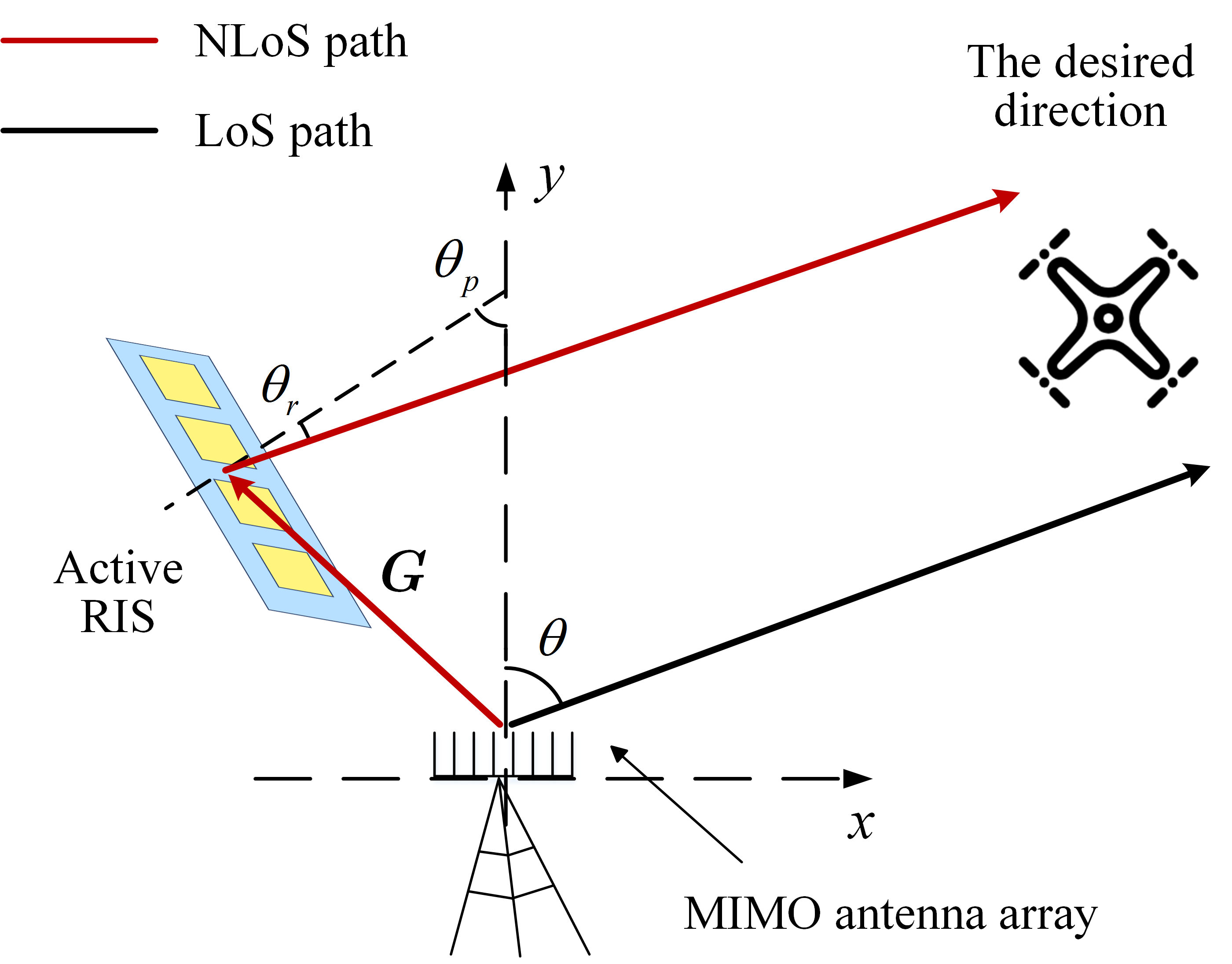}
	\caption{The ARIS-aided MIMO transmit array}
	\vspace{-6mm}
\end{figure}

\section{Signal Model and Problem Formulation}

\subsection{Signal Model}
Let us consider an ARIS-aided MIMO radar consisting of a colocated MIMO radar transceiver and a closely-spaced ARIS, as displayed in Fig. 1.
The MIMO radar equips with a $L_1$-element uniform linear array (ULA) as the transmit and receive antennas, while the ARIS is composed of $L_2$ active RIS elements to assist the transmit beampattern synthesis.
For a far-field potential target at the direction of departure (DoD) $\theta$ measured from the array broadside, there exist two paths, the LoS path and the NLoS path, from the MIMO array to the target direction.
As seen in Fig. 1, the LoS path directly transmits the probing signal to the target direction, while the NLoS path reflects it to the target direction via the ARIS.

Let the $l$-th element of MIMO transmit array radiate a different waveform $x_l(n)$ with $l=1, ...,L_1$ and $n=1, ...,N$, where $N$ is the sample number of each pulse.
Let $\bm{x}(n)=[x_1(n),x_2(n), \cdots, x_{L_1}(n)]^T \in \mathbb{C}^{L_1}$ be the space transmit waveform at the $n$-th sample and $\bm{X}=[\bm{x}(1), \bm{x}(2), \cdots, \bm{x}(N)] \in \mathbb{C}^{L_1 \times N}$ be the space-time waveform matrix.
Under the assumption of narrowband probing signals and non-dispersive propagation, for the LoS path, the $N$ samples of synthesized signal at the direction $\theta$ is 
\begin{equation}
    \bm{y}_{\rm{LoS}}=(\bm{a}^H(\theta)\bm{X})^T=(\bm{I}_{N}\otimes\bm{a}^H(\theta))\bm{x},
\end{equation}
where $\bm{x}=\mathrm{vec}(\bm{X})$ and $\bm{a}(\theta)$ is the transmit response vector of MIMO array given by
\begin{equation}
    \bm{a}(\theta) = [1,e^{\frac{j2\pi d \sin\theta}{\lambda}}, \cdots, e^{\frac{j2\pi (L_1-1)d \sin\theta}{\lambda}}]^T \in\mathbb{C}^{L_1},
\end{equation}
in which $\lambda$ is the wavelength of probing signals and $d$ is the inter-element space.

Since the ARIS is deployed close to the MIMO array, the MIMO array and ARIS have different DoDs for the same target \cite{Chen2024Reconfigurable}.
Denote $\theta_r$ as the target's DoD measured by the ARIS and $\theta_p$ as the angle between ARIS and MIMO array.
Defining the clockwise angle from the normal as negative, we then have $\theta_r=\theta+\theta_p$.
In this case, the transmit response vector of ARIS towards the direction $\theta_r$ is given by
\begin{equation}
{\bm b}(\theta_r) = [1, {e^{\frac{j2\pi {\tilde{d}}{\sin\theta_r}}{\lambda}}},...,{e^{{\frac{j2\pi (L_2-1){\tilde{d}}\sin\theta_r}{\lambda}}}}]^{T} \in \mathbb{C}^{L_2},
\end{equation}
where $\tilde{d}$ is the inter-element space of ARIS.
Generally, the ARIS is jointly described by an amplitude amplification matrix $\bm{P}\triangleq \mathrm{Diag}([p_1,p_2,..,p_{L_2}])$ and a phase shift matrix ${\bm\Phi}\triangleq \mathrm{Diag}([e^{j\phi_1},e^{j\phi_2},...,e^{j\phi_{L_2}}])$, where $p_l \geq 0$ and $\phi_l \in[0,2\pi]$ for $l=1,\cdots L_2$.
For simplicity, we define ${\bm V}\triangleq{\bm P}{\bm \Phi}$ as the reflection coefficients matrix and its diagonal element vector as $\bm{v} \triangleq{\rm diag}({\bm V}) =[v_1,v_2,\cdots,v_{L_2}]^T$ with $v_l=p_l e^{j\phi_l}$.

For a narrowband MIMO radar, its range resolution is often greater than the distance between ARIS and MIMO array due to close deployment.
We therefore assume that the illustrating signals from LoS and NLoS paths occupy the same range cell at all far-field desired locations.
Denote the channel matrix from MIMO array to ARIS as ${\bm G}\in \mathbb{C}^{L_2 \times L_1}$.
The $N$ samples of the synthesized signal at the NLoS path~is
\begin{equation}
    \bm{y}_{\rm{NLoS}}=(\bm{b}^H (\theta+\theta_p)\bm{VG}\bm{X})^T
    =(\bm{I}_{N}\otimes(\bm{b}^H(\theta+\theta_p)\bm{VG}))\bm{x}.
\end{equation}
Then the transmit beampattern $P(\theta)$ of the proposed ARIS-aided MIMO radar at the direction $\theta$ can be expressed as
\begin{equation}
	\label{beampattern}
	\begin{aligned}
	P(\theta) =(\bm{y}_{\rm{LoS}}+\bm{y}_{\rm{NLoS}})^H (\bm{y}_{\rm{LoS}}+\bm{y}_{\rm{NLoS}}) =\bm{x}^H \bm{R}(\theta) \bm{x},
	\end{aligned}
\end{equation}
where $\bm{R}(\theta)$ is defined by
\begin{equation}
\label{Rtheta}
    \bm{R}(\theta) =(\bm{I}_{N}\otimes\bm{c}^H(\theta))^H(\bm{I}_{N}\otimes\bm{c}^H(\theta)) =\bm{I}_{N}\otimes(\bm{c}(\theta)\bm{c}^H(\theta)),
\end{equation}
with $\bm{c}(\theta)=\bm{a}(\theta)+\bm{G}^H \bm{V}^H \bm{b}(\theta+\theta_p)$.
In practice, $\bm{G}$ can be effectively estimated by many existing approaches
\cite{Swindlehurst2022Channel,An2022Low}. Hence we assume it is known in advance.

In general, the whole direction region is divided into two sectors, the mainlobe region $\Theta_m$ and the sidelobe region $\Theta_s$, where $\Theta_m$ is usually composed of a continuous region for single transmit beam and several sub-regions for multiple transmit beams, and $\Theta_s$ consists of the remaining region.
To guarantee a good probing performance, the transmit energy is expected to be concentrated into the mainlobe region $\Theta_m$ as much as possible.
Therefore, the ISMR of transmit beampattern is an effective metric to quantify the degree of energy concentration \cite{Xu2015Colocated}.
Specifically, the ISMR of \eqref{beampattern} is defined as
\begin{equation}
\label{ISMR}
    \mathrm{ISMR} =\frac{\int_{\Theta_s}\bm{x}^H \bm{R}(\theta) \bm{x} d\theta} {\int_{\Theta_m} \bm{x}^H \bm{R}(\theta) \bm{x} d\theta}
    =\frac{\bm{x}^H \bm{\Omega}_s \bm{x}} {\bm{x}^H \bm{\Omega}_m \bm{x}}
\end{equation}
where $\bm{\Omega}_s=\bm{I}_{N}\otimes\bm{\Sigma}(\Theta_s)$, $\bm{\Omega}_m=\bm{I}_{N}\otimes\bm{\Sigma}(\Theta_m)$, and the matrix $\bm{\Sigma}(\Theta)$ is defined as
\begin{equation}
\label{SigmaTheta}
\begin{aligned}
    \bm{\Sigma}(\Theta) = \int_{\Theta} \bm{c}(\theta)\bm{c}^H(\theta) d\theta
     =\bm{A}_{\Theta} + \bm{G}^H \bm{V}^H \bm{D}_{\Theta} + \bm{D}_{\Theta}^H \bm{V}\bm{G} +\bm{G}^H \bm{V}^H \bm{B}_{\Theta} \bm{V}\bm{G},
\end{aligned}
\end{equation}
in which the integrals $\bm{A}_{\Theta}=\int_{\Theta}\bm{a}(\theta)\bm{a}^H(\theta) d\theta$, $\bm{D}_{\Theta}=\int_{\Theta}\bm{b}(\theta+\theta_p)\bm{a}^H(\theta) d\theta$ and $\bm{B}_{\Theta}=\int_{\Theta}\bm{b}(\theta+\theta_p)\bm{b}^H(\theta+\theta_p) d\theta$ over $\Theta$ (either $\Theta_m$ or $\Theta_s$) can be numerically calculated.
In fact, the integrals in $\bm{A}_{\Theta}$ and $\bm{B}_{\Theta}$ have analytically results if we use the normalized direction $\vartheta=\sin\theta$ \cite{Cheng2018Communication}.

\subsection{Practical Waveform Constraints}
Due to the limitation of transmitter hardware, such as power amplifiers and digital-to-analog converters, there exist several kinds of practical waveform constraints, involving CM \cite{Wang2012On}, PAR \cite{Stoica2008Waveform} and BM \cite{Yu2019Wideband}.

\subsubsection{CM Constraint}
~\\
\indent The CM constraint is often used to maximize the transmit efficiency so that the power amplifiers can operate in the saturation mode \cite{Tang2016Joint}.
This constraint is expressed as
\begin{equation}
\label{CMconstraint}
    \vert x_n \vert = \varepsilon_1, \ n=1,\cdots,NL_1,
\end{equation}
where $\varepsilon_1>0$ is the modulus of the waveform vector $\bm{x}$.
This constraint is denoted as $\mathcal{C}_1$ for short.

\subsubsection{PAR Constraint}
~\\
\indent As the CM constraint leads to a considerable performance loss on MIMO radars, such as receive SINR, it is desirable to relax the CM constraint and replace it with the PAR constraint as
\begin{equation}
\text{PAR}(\bm{x}) = \frac{\text{max}_n \ \vert x_n \vert^2}{\Vert\bm{x}\Vert_2^2/NL_1} \leq \eta, \ \eta \in [1,NL_1].
\end{equation}
where $\eta$ denotes the maximum allowable PAR \cite{Li2023Multispectrally}.
Generally, the PAR constraint can be rewritten as
\begin{equation}
\label{PARconstraint}
\Vert\bm{x}\Vert_2^2=E, \ \vert x_n \vert \leq \sqrt{\frac{\eta E}{NL_1}}, \ n=1,\cdots,NL_1,
\end{equation}
where $E$ is the total transmit energy.
When $\eta=1$, the PAR constraint degenerates into the CM constraint.
This constraint is denoted as $\mathcal{C}_2$ for short.

\subsubsection{BM Constraint}
~\\
\indent Even though the PAR constraint make a restriction on the maximum amplitude of transmit waveforms, it may result in a high dynamic range, which is prohibitive in practical hardware, such as digital-to-analog converters and linear power amplifiers.
Based on this consideration, the BM constraint is an alternative relaxation of the CM constraint.
Specifically, the BM constraint is represented as
\begin{equation}
\label{BMconstraint}
   {\varepsilon_2-\delta}\leq \vert x_n \vert \leq  {\varepsilon_2+\delta}, \ n=1,\cdots,NL_1,
\end{equation}
where ${\varepsilon_2+\delta}$ and ${\varepsilon_2-\delta}$ are upper and lower modulus bounds, respectively, and $0 \leq \delta \leq \epsilon_2$.
The upper and lower bounds enforced on waveform amplitudes explicitly limit the dynamic range.
The PAR constraint with a given total transmit energy $E$ and the CM constraint can be respectively obtained by setting $\varepsilon_2=\delta= \sqrt{\frac{\eta E}{4NL_1}}$ and $ \varepsilon_2=\varepsilon_1$ with $\delta=0$ \cite{Yu2019Wideband}.
This constraint is denoted as $\mathcal{C}_3$ for short.

%

For convenience, the waveform constraints are uniformly expressed by
\begin{equation}
    \mathcal{X} = \{\bm{x} \in \mathbb{C}^{NL_1} \} \cap (\cap_i \mathcal{C}_i).
\end{equation}
Note that the waveform parameters, i.e., the waveform amplitude $\varepsilon_1$ in the CM constraint \eqref{CMconstraint}, the total transmit energy $E$ in the PAR constraint \eqref{PARconstraint} and the parameter $\varepsilon_2$ in the BM constraint \eqref{BMconstraint}, have no effect on the ISMR in \eqref{ISMR}.
Without loss of generality, we set $\varepsilon_1=1$, $E=NL_1$ and $\varepsilon_2=1$, respectively.


\subsection{Problem Formulation}
For the ARIS-aided MIMO radar, the proposed beampattern synthesis is to minimize the ISMR by jointly designing the transmit waveform $\bm{x}$ and the ARIS reflection coefficients $\bm{V}$ ($\bm{v}$) under practical constraints.

As an ARIS has the ability to amplify the incident signal, it inevitably introduces a non-ignorable noise into the reflected signal \cite{Zhang2023Active}.
Due to the finite power, the reflecting power of ARIS is confined by
\begin{equation}
\label{RISpower}
    \Vert {\bm V}{\bm G}{\bm x} (n) \Vert^2_2+\Vert {\bm V} \Vert^2_{\text{F}}\sigma_v^2\leq P_\mathrm{A}, \ n=1,\cdots,N.
\end{equation}
where $\sigma^2_v$ is the incident noise variance and $P_\mathrm{A}$ is the maximum available power.
On the other hand, the ARIS element has a finite amplification gain since it only equips with simple amplifying circuits \cite{Long2021Active}.
Then the ARIS reflection coefficients are restricted by
\begin{equation}
\label{RISamplification}
    |v_l| \leq \varsigma, \ l=1,...,L_2,
\end{equation}
where $\varsigma$ is the allowable maximum amplification factor.

In combination with practical waveform and ARIS constraints, we formulate the proposed beampattern synthesis into the following optimization problem,
\begin{equation}
\label{Prob1}
\begin{split}
    \min_{{\bm x}, {\bm V}} \ \frac{\bm{x}^H \bm{\Omega}_s \bm{x}} {\bm{x}^H \bm{\Omega}_m \bm{x}} \quad
    {\rm{s.t.}}\ &\Vert {\bm V}{\bm G}{\bm x} (n) \Vert^2_2+\Vert {\bm V} \Vert^2_{\text{F}}\sigma_v^2\leq P_\mathrm{A}, \ \forall n.\\
    & |v_l| \leq \varsigma, \ \forall l, \\
    & \bm{x} \in \mathcal{X}.
\end{split}
\end{equation}
Note that \eqref{Prob1} is a complicated nonconvex constrained FP, because the objective function $\varGamma(\bm{x},\bm{V}) \triangleq \frac{\bm{x}^H\bm{\Omega}_s\bm{x}} {\bm{x}^H\bm{\Omega}_m\bm{x}}$ is nonconvex over the variables $\bm{x}$ and $\bm{V}$, the waveform constraints form a nonconvex feasible set and the variables $\bm{x}$ and $\bm{V}$ are coupled in the ARIS power constraints and $\varGamma(\bm{x},\bm{V})$.
It is intractable to directly solve the problem \eqref{Prob1}.
In the next section, we will derive an equivalent formulation of \eqref{Prob1} and then create an effective algorithm to handle with this challenging problem.

\section{Proposed Beampattern Synthesis Algorithm for ARIS-Aided MIMO Radar}
In this section, we provide an AM-based algorithm to tackle the problem \eqref{Prob1} by leveraging the FP, CADMM and CCCP schemes. 
As the objective function $\varGamma(\bm{x},\bm{V})$ in \eqref{Prob1} has a biquadratic fractional expression, it is difficult to be processed directly.
We first adopt Dinkelbach transform \cite{Tang2020Polyphase} to convert it into an equivalent integral form as
\begin{equation}
\label{Objfun1}
    \varUpsilon(\bm{x},\bm{V},\mu) \triangleq \mu\bm{x}^H\bm{\Omega}_s\bm{x} -\bm{x}^H\bm{\Omega}_m\bm{x},
\end{equation}
where $\mu=\frac{\bm{x}^H\bm{\Omega}_m\bm{x}} {\bm{x}^H\bm{\Omega}_s\bm{x}}=\frac{1}{\varGamma(\bm{x},\bm{V})}$ is an auxiliary variable.
By alternately minimizing $\varUpsilon(\bm{x,V},\mu)$ and updating the variable $\mu$, the
minimization of $\varGamma(\bm{x,V})$ can converge to a locally optimal solution.
As a result, we convert the problem \eqref{Prob1}~into
\begin{equation}
\label{Prob2}
\begin{split}
\min_{\bm{x}, \bm{V}} \ & \varUpsilon(\bm{x},\bm{V},\mu)  \\
{\rm{s.t.}}\ &\Vert {\bm V}{\bm G}{\bm x} (n) \Vert^2_2+\Vert {\bm V} \Vert^2_{\text{F}}\sigma_v^2\leq P_\mathrm{A}, \ \forall n,\\
& |v_l| \leq \varsigma, \ \forall l, \\
& \bm{x} \in \mathcal{X}.
\end{split}
\end{equation}
Note that the variables $\bm{x}$ and $\bm{V}$ are still coupled in $\varUpsilon(\bm{x},\bm{V},\mu)$ and the ARIS power constraints.
We optimize these variables separately based on the AM framework as follows.

\vspace{-2mm}
\subsection{Optimize $\bm{x}$ with fixed $\bm{V}$ and $\mu$}
In this subsection, we fix the variables $\bm{V}$ and $\mu$ and optimize the transmit waveform $\bm{x}$.
Removing the irrelative items, we rewrite \eqref{Prob2} over the variable $\bm{x}$ at the $t$-th iterations~as
\begin{subequations}
\label{Probx1}
	\begin{align}
	\label{Probx11}
    \min_{\bm{x}} &\ \bm{x}^H(\mu^t\bm{\Omega}^t_s-\bm{\Omega}^t_m)\bm{x}, \\
	\label{Probx12}
    {\rm{s.t.}} &\ \Vert {\bm V}^t{\bm G}{\bm x} (n) \Vert^2_2+\Vert {\bm V}^t \Vert^2_{\text{F}}\sigma_v^2\leq P_\mathrm{A}, \ \forall n,\\
	\label{Probx13}
	& \bm{x} \in \mathcal{X}.
	\end{align}
\end{subequations}
Note that \eqref{Probx12} is convex and separate for the waveform ${\bm x}$, while \eqref{Probx13} is nonconvex and sometimes inseparable.
It is intractable to tackle these constraints simultaneously.
We thus divide and rule the constraints as follows.

Define the auxiliary variables satisfying $\bar{\bm x} = {\bm x}$ and $\hat{\bm x} = {\bm x}$.
According to the CADMM principle at the scaled-form \cite{Boyd2010distributed}, we convert the problem \eqref{Probx1} into
\begin{subequations}
	\label{ADMMprob1}
	\begin{align}
	\label{ADMMsub11a}
	\min_{{\bm x},\bar{\bm x},\hat{\bm x},\bm{p},\bm{r}} & \bm{x}^H(\mu^t\bm{\Omega}^t_s-\bm{\Omega}^t_m)\bm{x} +\frac{\beta_1}{2}(\Vert \bar{\bm x}-\bm{x}+\bm{p} \Vert_2^2 -\Vert \bm{p} \Vert_2^2)	
	+\frac{\beta_1}{2}(\Vert \hat{\bm x}-\bm{x}+\bm{r} \Vert_2^2 -\Vert \bm{r} \Vert_2^2) \\
	\label{ADMMsub12a}
	{\rm{s.t.}} \enspace &\Vert {\bm V}^t {\bm G}\bar{\bm x}(n) \Vert^2_2+\Vert {\bm V}^t \Vert^2_{\text{F}}\sigma_v^2\leq P_\mathrm{A}, \ \forall n, \\
	\label{ADMMsub13a}
	& \hat{\bm{x}} \in \mathcal{X}(\bm{\chi}),
	\end{align}
\end{subequations}
where $\beta_1>0$ is the penalty parameter, $\bm{p}$ and $\bm{r}$ are the scaled dual variables.
At the $(k+1)$-th iteration, the CADMM algorithm consists of the following update procedures:
\begin{subequations}
	\begin{align}
	\label{ADMM:sub1a}	
	\bar{\bm{x}}^{k+1}&=\arg\min_{\bar{\bm{x}}} \Vert \bar{\bm x}-\bm{x}^k+\bm{p}^k \Vert_2^2 \quad {\rm{s.t.}} \enspace \Vert {\bm V}^t {\bm G}\bar{\bm x}(n) \Vert^2_2+\Vert {\bm V}^t \Vert^2_{\text{F}}\sigma_v^2\leq P_\mathrm{A}, \ \forall n.   \\
	\label{ADMM:sub2a}
	\hat{\bm{x}}^{k+1}&= \arg \min_{\hat{\bm{x}},\bm{\chi}} \Vert \hat{\bm x}-\bm{x}^k+\bm{r}^k \Vert_2^2 	\quad {\rm{s.t.}} \enspace \hat{\bm{x}} \in \mathcal{X}(\bm{\chi}), \\
	\label{ADMM:sub3a}
	\bm{x}^{k+1}&=\arg\min_{\bm{x}} \ \bm{x}^H(\mu^t\bm{\Omega}^t_s-\bm{\Omega}^t_m)\bm{x} +\frac{\beta_1}{2}\Vert \bar{\bm x}^{k+1}-\bm{x}+\bm{p}^{k} \Vert_2^2  +\frac{\beta_1}{2}\Vert \hat{\bm x}^{k+1}-\bm{x}+\bm{r}^{k} \Vert_2^2,	\\	
	\label{ADMM:sub4a}
	{\bm p}^{k+1}&={\bm p}^{k}+\bar{\bm x}^{k+1}-{\bm x}^{k+1}, \\
	\label{ADMM:sub5a}
	{\bm r}^{k+1}&={\bm r}^{k}+\hat{\bm x}^{k+1}-{\bm x}^{k+1}.
	\end{align}
\end{subequations}

\subsubsection{Update $\bar{\bm{x}}^{k+1}$ with given $\{{\bm x}^k,\hat{\bm x}^k,\bm{p}^k,\bm{r}^k\}$}
~\\
\indent The problem \eqref{ADMM:sub1a} can be divided into $N$ simple convex QCQP with one constraint (QCQP-1) problems, and the $n$-th subproblem is expressed as
\begin{equation}
\label{ADMMsub1a}
\begin{aligned}	
    \min_{\bar{\bm{x}}(n)}  \Vert \bar{\bm x}(n)-\bm{x}^k(n)+\bm{p}^k(n) \Vert_2^2 \quad
    {\rm{s.t.}} \enspace \bar{\bm x}^H(n){\bm R}\bar{\bm x}(n) \leq \bar{P}_\mathrm{A},
\end{aligned}
\end{equation}
where $\bm{R}={\bm G}^H({\bm V}^t)^H {\bm V}^t{\bm G}$ and $\bar{P}_\mathrm{A}=P_\mathrm{A}-\Vert {\bm V}^t \Vert^2_{\text{F}}\sigma_v^2$.
Here, we solve the problem \eqref{ADMMsub1a} efficiently by using the Lagrangian dual method.

Define the Lagrangian function of \eqref{ADMMsub1a} as
\begin{equation}
\label{LagrangeSub1a}
    L(\bar{\bm x}(n), \varepsilon) =\Vert \bar{\bm x}(n)-\tilde{\bm{x}}(n) \Vert_2^2 + \varepsilon (\bar{\bm x}^H(n){\bm R}\bar{\bm x}(n) -\bar{P}_\mathrm{A}).
\end{equation}
where $\tilde{\bm{x}}(n)=\bm{x}^k(n)-\bm{p}^k(n)$.
As $L(\bar{\bm x}(n), \varepsilon)$ is a convex quadratic function of $\bar{\bm x}(n)$, we can find the minimizing $\bar{\bm x}(n)$ from the first-order optimality condition
\begin{equation}
\label{OptimalxnCond}
    \nabla_{\bar{\bm x}^{*}(n)} L(\bar{\bm x}(n),\varepsilon) =\bar{\bm x}(n) +\varepsilon\bm{R}\bar{\bm x}(n) -\tilde{\bm{x}}(n) = 0,
\end{equation}
which yields
\begin{equation}
\label{Optimalxbn0}
    \bar{\bm x}(n) =({\bm I}_{L_1} +\varepsilon\bm{R})^{-1}\tilde{\bm{x}}(n).
\end{equation}
Inserting \eqref{Optimalxbn0} into \eqref{LagrangeSub1a}, we have the dual function as
\begin{equation}
\label{Dualfunction0}
    g(\varepsilon) = -\tilde{\bm{x}}^H(n)({\bm I}_{L_1} +\varepsilon\bm{R})^{-1}\tilde{\bm{x}}(n)  +\tilde{\bm{x}}^H(n)\tilde{\bm{x}}(n) -\varepsilon \bar{P}_\mathrm{A}.
\end{equation}
Therefore, the Lagrange dual problem of \eqref{ADMMsub1a} is
\begin{equation}
\label{DualProb1}
    \min_{\varepsilon} \ \tilde{\bm{x}}^H(n)({\bm I}_{L_1}+\varepsilon\bm{R})^{-1} \tilde{\bm{x}}(n) +\varepsilon \bar{P}_\mathrm{A}, \ {\rm{s.t.}} \ \varepsilon \geq 0.
\end{equation}

Define the eigenvalue decomposition of $\bm{R}$ as ${\bm{R}} ={\bm{T}} {\bm{\Gamma}} {\bm{T}}^H$, where $\bm{T}$ is the eigenvector matrix, ${\bm{\Gamma}}=\text{diag}([\gamma_1,...,\gamma_{L_1}])$ is the eigenvalue matrix, and the eigenvalues obey $\gamma_1 \geq \gamma_2 \geq... \geq \gamma_{L_1} \geq 0$.
Inserting ${\bm{R}} = {\bm{T}} {\bm{\Gamma}} {\bm{T}}^H$ into \eqref{DualProb1}, we have the equivalent expression of \eqref{DualProb1} as
\begin{equation}
\label{DualProb11}
    \min_{\varepsilon} \ \sum_{l=1}^{L_1}\frac{{|\tilde{x}_l^{\prime}|}^2}{1+\varepsilon\gamma_l} +\varepsilon\bar{P}_\mathrm{A}, \enspace {\rm{s.t.}} \enspace \varepsilon \geq 0.
\end{equation}
where ${\bm{T}}^H \tilde{\bm{x}}(n)=[\tilde{x}_1^{\prime}, \tilde{x}_2^{\prime} ,...,\tilde{x}_{L_1}^{\prime}]^T$.
Then, we calculate the first-order derivative of the objective function in \eqref{DualProb11} as
\begin{equation}
\label{Multiplier1}
    \zeta(\varepsilon) = -\sum_{l=1}^{L_1} \frac{\gamma_l{|\tilde{x}_l^{\prime}|}^2}{(1+\varepsilon\gamma_l)^2} +\bar{P}_\mathrm{A}.
\end{equation}
Obviously, $\zeta(\varepsilon)$ is monotonically increasing with $\varepsilon \geq 0$.
We set $\zeta(\varepsilon)=0$ to find the optimal solution of \eqref{DualProb11}.
If $\sum_{l=1}^{L_1}\gamma_l{|\tilde{x}_l^{\prime}|}^2 < \bar{P}_\mathrm{A}$, $\zeta(\varepsilon)=0$ always does not hold.
In this case, the solution of \eqref{DualProb11} is $\varepsilon^{\star} = 0$, which implies that the constraint in \eqref{ADMMsub1a} is not active.
Otherwise, the solution $\varepsilon^{\star}$ can be easily obtained by solving this nonlinear equation, via for example bisection or Newton’s method.

Inserting $\varepsilon^{\star}$ into \eqref{Optimalxbn0}, we have the optimal solution of \eqref{ADMMsub1a} as
\begin{equation}
\label{solutionxbn}
\bar{\bm x}^{k+1}(n) =({\bm I}_{L_1} +\varepsilon^{\star}\bm{R})^{-1}\tilde{\bm{x}}(n).
\end{equation}

\subsubsection{Update $\hat{\bm{x}}^{k+1}$ with given $\{{\bm x}^k,\bar{\bm x}^{k+1},\bm{p}^k,\bm{r}^k\}$}
~\\
\indent
For each waveform constraint \eqref{CMconstraint}, \eqref{PARconstraint} and \eqref{BMconstraint}, \eqref{ADMM:sub2a} is a nonconvex constrained optimization problem.
In the following, we give the solution of \eqref{ADMM:sub2a} with each waveform constraint, respectively.

For the CM constraint, \eqref{ADMM:sub2a} is expressed as
\begin{equation}
\label{ADMMsub2aCM}
    \hat{\bm{x}}^{k+1}= \arg \min_{\hat{\bm{x}}} \Vert \hat{\bm x}-\bm{x}^k+\bm{r}^k \Vert_2^2 	\quad {\rm{s.t.}} \enspace \vert \hat{x}_n \vert =1, \ \forall n.
\end{equation}
The problem \eqref{ADMMsub2aCM} can be divided into the following problems
\begin{equation}
\label{sub2aCM}
    \min_{\hat{x}_n} \ |\hat{x}_n-x^k_n+r^k_n|^2 \ \ \ {\rm{s.t.}}\ \vert\hat{x}_n\vert=1,
\end{equation}
for each $n=1,\cdots,NL_1$. Then the solution of \eqref{ADMMsub2aCM} is
\begin{equation}
\label{solxhnCM}
    \hat{x}^{k+1}_n =\frac{x^k_n-r^k_n}{\vert x^k_n-r^k_n \vert},\ \forall n.
\end{equation}

For the PAR constraint, \eqref{ADMM:sub2a} is described as
\begin{equation}
\label{ADMMsub2aPAR}
\begin{aligned}	
    \hat{\bm{x}}^{k+1}= \arg \min_{\hat{\bm{x}}} \Vert \hat{\bm x}-\bm{x}^k+\bm{r}^k \Vert_2^2 	
    \quad {\rm{s.t.}} \enspace \Vert\hat{\bm{x}}\Vert_2^2=NL_1, \ \vert \hat{x}_n \vert \leq \sqrt{\eta}, \ \forall n.
\end{aligned}	
\end{equation}
Note that \eqref{ADMMsub2aPAR} aims to find the nearest vector of $\tilde{\bm{x}}^k=\bm{x}^k-\bm{r}^k$ with the PAR constraint.
Its update solution is readily derived by using the Karush-Kuhn-Tucker (KKT) condition provided in \cite{Tropp2005Designing}.
The solving procedure for \eqref{ADMMsub2aPAR} is summarized in Algorithm 1.

\begin{figure}[!t]
	\label{alg:PARNest}
	\begin{algorithm}[H]
		\caption{Solution to \eqref{ADMMsub2aPAR}.}
		\begin{algorithmic}[1]
			\renewcommand{\algorithmicrequire}{\textbf{Input:}}
			\REQUIRE $\tilde{\bm{x}}^k$, $\eta$. \\         
			\renewcommand{\algorithmicrequire}{\textbf{Initialization:}}
			\REQUIRE Scale $\tilde{\bm{x}}^k$ to have unit norm and set $n=0$. \\ 
			\WHILE {$n \leq NL_1$}
			\STATE Let the set $\mathcal{N}$ index $(NL_1-n)$ components of $\tilde{\bm{x}}^k$ with least magnitude. \\
			If this set is not uniquely determined, increment $n$ and repeat Step 1.
			\STATE If $\tilde{x}^k_n=0$ for each $n$ in $\mathcal{N}$, obtain the optimal solution as \\
			\begin{equation}
				\begin{aligned}
					\hat{x}^{k+1}_n=\left\{
					\begin{array}{rcl}
						\sqrt{\frac{NL_1-n\eta}{NL_1-n}}\angle{\tilde{x}^k_n},&   &{\text{if}}\ n\in\mathcal{N}  \\
						\sqrt{\eta}\angle{\tilde{x}^k_n},&   &{\text{if}}\ n\not\in\mathcal{N} \notag
					\end{array}\right.
				\end{aligned}
			\end{equation}
			Otherwise, calculate \\
			\vspace{2mm}
			$\qquad \qquad \quad$	$\eta_1=\sqrt{\frac{NL_1-n\eta}{\sum\limits_{n\in\mathcal{N}} \vert\tilde{x}^k_n\vert^2}}$
			\vspace{2mm}
			\STATE If $\vert\tilde{x}^k_n\vert> \frac{\sqrt{\eta}}{\eta_1}$ for any $n\in\mathcal{N}$, increment $n$ and return to Step 1; \\
			Otherwise, break the while.
			\ENDWHILE
			\renewcommand{\algorithmicensure}{\textbf{Output:}}
			\ENSURE Optimal solution of $\hat{\bm{x}}^{k+1}$ as \\
			\begin{equation}
				\begin{aligned}
					\hat{x}^{k+1}_n=\left\{
					\begin{array}{rcl}
						\eta_1\tilde{x}^k_n,&   &{\text{if}}\ n\in\mathcal{N}  \\
						\sqrt{\eta}\angle{\tilde{x}^k_n},&   &{\text{if}}\ n\not\in\mathcal{N} \notag
					\end{array}\right.
				\end{aligned}
			\end{equation}    
		\end{algorithmic}
	\end{algorithm}
	\vspace{-10mm}
\end{figure}

For the BM constraint, \eqref{ADMM:sub2a} is described as
\begin{equation}
\label{ADMMsub2cBM}
\begin{aligned}	
    \hat{\bm{x}}^{k+1}= \arg \min_{\hat{\bm{x}}} \Vert \hat{\bm x}-\bm{x}^k+\bm{r}^k \Vert_2^2 
    \quad {\rm{s.t.}} \enspace {1-\delta} \leq \vert \hat{x}_n \vert \leq 1+\delta, \ \forall n.
\end{aligned}	
\end{equation}
Similar to the CM constraint, $\hat{x}_n$ is determined by solving the following problems
\begin{equation}
\label{sub2cBM}
\min_{\hat{x}_n} \ |\hat{x}_n-x^k_n+r^k_n|^2 \ \
 {\rm{s.t.}}  \ {1-\delta} \leq \vert \hat{x}_n \vert \leq {1+\delta}, \ \forall n.
\end{equation}
And the solution of \eqref{sub2cBM} is easily given by 
\begin{equation}
\label{solxhnBM}	
\begin{aligned}
	\hat{x}^{k+1}_n=\left\{
	\begin{array}{rcl}
		(1-\delta) \angle{(x^k_n-r^k_n)},&  &{\text{if}}\ |x^k_n-r^k_n| < 1-\delta \\
		(1+\delta) \angle{(x^k_n-r^k_n)},&  &{\text{if}}\ |x^k_n-r^k_n| > 1+\delta \\
		{x^k_n-r^k_n},&  &\text{otherwise}.
	\end{array}\right.
\end{aligned}
\end{equation}


\subsubsection{Update $\bm{x}^{k+1}$ with given $\{\bar{\bm x}^{k+1},\hat{\bm{x}}^{k+1},\bm{p}^k,\bm{r}^k\}$}
~\\
\indent The problem \eqref{ADMM:sub3a} is a unconstrained and convex quadratic program problem, and its global optimal solution is
\begin{equation}
\begin{aligned}
\label{solxn}
    \bm{x}^{k+1} =\frac{\beta_1}{2}(\beta_1\bm{I}_{NL_1} +\mu^t\bm{\Omega}^t_s-\bm{\Omega}^t_m)^{-1}(\bar{\bm x}^{k+1}+\hat{\bm{x}}^{k+1}+\bm{p}^{k}+\bm{r}^{k}).
\end{aligned}	
\end{equation}

Based on the above discussions, the solving procedure for \eqref{ADMMprob1} is summarized in Algorithm 2, where $\iota_1$ is the termination tolerance and $K_{\max}$ is the maximum iteration number.

\begin{figure}[!t]
	\label{alg:CADMM}
	\begin{algorithm}[H]
		\caption{CADMM-based Algorithm for Solving \eqref{ADMMprob1}.}
		\begin{algorithmic}[1]
			\renewcommand{\algorithmicrequire}{\textbf{Input:}}
			\REQUIRE $\mu^{t}$, $\bm{\Omega}^t_s$, $\bm{\Omega}^t_m$, $\bm{G}$, $P_A$, $\sigma_v^2$, $\iota_1$ and $K_{\max}$. \\         
			\renewcommand{\algorithmicrequire}{\textbf{Initialization:}}
			\REQUIRE Set $k=0$, initialize $\bm{p}^{0}$ and $\bm{r}^{0}$.  
			\WHILE {$k \leq K_{\max}$ and for any $\Vert\bm{p}^{k+1}-\bm{p}^{k}\Vert_2>\iota_1$ or $\Vert\bm{r}^{k+1}-\bm{r}^{k}\Vert_2>\iota_1$}
			\STATE Obtain $\bar{\bm x}^{k+1}$ using \eqref{solutionxbn};
			\STATE Obtain $\hat{\bm{x}}^{k+1}$ using \eqref{solxhnCM} for CM constraint, Algorithm 1 for PAR constraint, and \eqref{solxhnBM} for BM constraint;
			\STATE Obtain $\bm{x}^{k+1}$ using \eqref{solxn};
			\STATE Update $\bm{p}^{k+1}$ and $\bm{r}^{k+1}$ using \eqref{ADMM:sub4a} and \eqref{ADMM:sub5a}, respectively;
			\STATE $k=k+1$;
			\ENDWHILE
			\renewcommand{\algorithmicensure}{\textbf{Output:}}
			\ENSURE MIMO transmit waveform $\bm{x}^{k+1}$.  
		\end{algorithmic}
	\end{algorithm}
	\vspace{-10mm}
\end{figure}

\vspace{-2mm}
\subsection{Optimize $\bm{V}$ with fixed $\bm{x}$ and $\mu$}
In this subsection, we fix the variables $\bm{x}$ and $\mu$ and optimize the ARIS reflection coefficients $\bm{V}$.
Before proceeding, we represent the function $\varUpsilon(\bm{x,V},\mu)$ as an explicit form over the variable $\bm{V}$ (or $\bm{v}$).

The beampattern $P(\theta)$ in \eqref{beampattern} can also be described as
\begin{equation}
\label{beampattern1}
    P(\theta)=\bm{c}^H(\theta) \bm{X} \bm{X}^H \bm{c}(\theta).
\end{equation}
Denoting $\bm{\Omega}=\bm{I}_{N}\otimes\bm{\Sigma}(\Theta)$, we find that
\begin{equation}
\begin{aligned}
\label{Intpattern1}
    \bm{x}^H \bm{\Omega} \bm{x}
    &=\mathrm{tr}\left({\int_{\Theta} \bm{c}^H(\theta) \bm{X} \bm{X}^H \bm{c}(\theta) d\theta} \right) 
    =\mathrm{tr}\left({\int_{\Theta} \bm{c}(\theta) \bm{c}^H(\theta) d\theta} \bm{X}\bm{X}^H \right)    \\
    &=\mathrm{tr}\left(\bm{\Sigma}(\Theta) \bm{X}\bm{X}^H \right).
\end{aligned}
\end{equation}

Inserting \eqref{SigmaTheta} into \eqref{Intpattern1}, we then have
\begin{equation}
\begin{aligned}
\label{Intpattern2}
     \bm{x}^H \bm{\Omega} \bm{x}
     =&\ \mathrm{tr}\left(\bm{A}_{\Theta}\bm{X}\bm{X}^H\right) +2\mathrm{Re}\{\mathrm{tr}\left(\bm{G}^H \bm{V}^H \bm{D}_{\Theta} \bm{X}\bm{X}^H\right)\} 
     +\mathrm{tr}\left(\bm{G}^H \bm{V}^H \bm{B}_{\Theta} \bm{V}\bm{G}\bm{X}\bm{X}^H\right) \\
     =&\ \mathrm{tr}\left(\bm{A}_{\Theta}\bm{X}\bm{X}^H\right) +2\mathrm{Re}\{\mathrm{tr}(\bm{V}^H \hat{\bm{D}}_{\Theta})\} 
     +\mathrm{tr}\left(\bm{V}^H \bm{B}_{\Theta} \bm{V} \bm{H} \right),
\end{aligned}
\end{equation}
where $\hat{\bm{D}}_{\Theta}=\bm{D}_{\Theta} \bm{X}\bm{X}^H \bm{G}^H$ and $\bm{H}=\bm{G}\bm{X}\bm{X}^H\bm{G}^H$.
According to the property of matrix trace, we have
\begin{equation}
\label{TrVB}
    \mathrm{tr}(\bm{V}^{H}\hat{\bm{D}}_{\Theta})=\bm{v}^{H}\hat{\bm{d}}_{\Theta},
\end{equation}
with $\hat{\bm{d}}_{\Theta}=\mathrm{diag}(\hat{\bm{D}}_{\Theta})$, and
\begin{equation}
\label{TrVBVC}
    \mathrm{tr}(\bm{V}^{H}\bm{B}_{\Theta}\bm{V}\bm{H})=\bm{v}^{H}(\bm{B}_{\Theta}\odot\bm{H}^{T})\bm{v}.
\end{equation}
Denoting $\hat{\bm{B}}_{\Theta}=\bm{B}_{\Theta}\odot\bm{H}^{T}$, we find that $\hat{\bm{B}}_{\Theta}$ is positive semi-definite since $\bm{B}_{\Theta}$ and $\bm{H}$ are both Hermitian positive semi-definite.

Substituting \eqref{TrVB} and \eqref{TrVBVC} into \eqref{Intpattern2}, we obtain
\begin{equation}
\label{Intpattern3}
    \bm{x}^H\bm{\Omega}\bm{x} =\mathrm{tr}\left(\bm{A}_{\Theta}\bm{X}\bm{X}^H\right) +2\mathrm{Re}\{\bm{v}^{H}\hat{\bm{d}}_{\Theta}\}
    +\bm{v}^{H}\hat{\bm{B}}_{\Theta}\bm{v},
\end{equation}
Therefore, we can rewrite $\varUpsilon(\bm{x,V},\mu)$ as
\begin{equation}
\begin{aligned}
\label{Objfun1v}
    \varUpsilon(\bm{x,V},\mu) =
    &\ \mu\mathrm{tr}\left(\bm{A}_{\Theta_s}\bm{X}\bm{X}^H\right) +2\mu\mathrm{Re}\{\bm{v}^{H} \hat{\bm{d}}_{\Theta_s}\} 
    +\mu\bm{v}^{H}\hat{\bm{B}}_{\Theta_s}\bm{v}     -\mathrm{tr}\left(\bm{A}_{\Theta_m}\bm{X}\bm{X}^H\right) \\ &-2\mathrm{Re}\{\bm{v}^{H}\hat{\bm{d}}_{\Theta_m}\}
    -\bm{v}^{H}\hat{\bm{B}}_{\Theta_m}\bm{v}
\end{aligned}
\end{equation}

On the other hand, we express the ARIS power constraint \eqref{RISpower} as
\begin{equation}
\label{RISpowerV}
    \bm{v}^H \bm{\Pi}_n \bm{v} \le P_\mathrm{A}, \ \forall n,
\end{equation}
where $\bm{\Pi}_n=\mathrm{Diag}(\bm{G}\bm{x}(n))^H \mathrm{Diag}(\bm{G}\bm{x}(n)) +\sigma^{2}_{v}\bm{I}_{L_2}$.

Substituting \eqref{Objfun1v} and \eqref{RISpowerV} into \eqref{Prob2} and discarding the irrelative items, we obtain the optimization problem over the variable $\bm{v}$ at the $t$-th iteration as
\begin{subequations}
	\label{ProbV1}
	\begin{align}
	\label{ProbV11}
	\min_{\bm{v}}
	&\ \mu^t\bm{v}^{H}\hat{\bm{B}}_{\Theta_s}^{t+1}\bm{v}
	-\bm{v}^{H}\hat{\bm{B}}_{\Theta_m}^{t+1}\bm{v}
	+2\mathrm{Re}\{\bm{v}^{H}\hat{\bm{d}}^{t+1}\}   \\
	\label{ProbV12}
	{\rm{s.t.}} &\  \bm{v}^H \bm{\Pi}_n^{t+1} \bm{v} \le P_A, \ \forall n.\\
	\label{ProbV13}
	&\ |v_l| \leq \varsigma, \ \forall l.
	\end{align}
\end{subequations}
where $\hat{\bm{d}}^{t+1}=\mu^t\hat{\bm{d}}_{\Theta_s}^{t+1}-\hat{\bm{d}}_{\Theta_m}^{t+1}$.
It is seen that the constraints \eqref{ProbV12} and \eqref{ProbV13} are both convex, while the objective function \eqref{ProbV11} is a difference of convex (DC) function which may be nonconvex.
Hence we resort to the CCCP to find its locally optimal solution \cite{2009GertCCP}.
Concretely, we replace the concave part of DC function with its first-order Taylor expansion and then iteratively solve a sequence of convex subproblems.


\begin{figure}[!t]
	\label{alg:CCP}
	\begin{algorithm}[H]
		\caption{CCCP-based Algorithm for Solving \eqref{ProbV1}.}
		\begin{algorithmic}[1]
			\renewcommand{\algorithmicrequire}{\textbf{Input:}}
			\REQUIRE $\mu^t$, $\hat{\bm{B}}_{\Theta_m}^{t+1}$, $\hat{\bm{B}}_{\Theta_s}^{t+1}$, $\hat{\bm{d}}^{t+1}$, $\bm{\Pi}_n^{t+1}$, 
            $\iota_2$ and $J_{\max}$. \\         
			\renewcommand{\algorithmicrequire}{\textbf{Initialization:}}
			\REQUIRE Set $j=0$, initialize $\bm{v}^{0}$.  
			\WHILE {$j \leq J_{\max}$ and $\Vert \bm{v}^{j+1}-\bm{v}^{j} \Vert_2>\iota_2$}
			\STATE Obtain $\bm{v}^{j+1}$ by solving the convex problem \eqref{ProbV2};
			\STATE $j=j+1$;
			\ENDWHILE
			\renewcommand{\algorithmicensure}{\textbf{Output:}}
			\ENSURE ARIS reflection coefficients $\bm{V}^{j+1}=\mathrm{Diag}(\bm{v}^{j+1})$.  
		\end{algorithmic}
	\end{algorithm}
	\vspace{-10mm}
\end{figure}

Expanding the function $\bm{v}^{H}\hat{\bm{B}}_{\Theta_m}^{t+1}\bm{v}$ at the point $\bm{v}^j$, we have 
\begin{align}
\label{MMFun}
    \bm{v}^{H}\hat{\bm{B}}_{\Theta_m}^{t+1}\bm{v} &\geq 2\mathrm{Re}\{\bm{v}^{H}\hat{\bm{B}}_{\Theta_m}^{t+1}\bm{v}^j\} -(\bm{v}^j)^{H}\hat{\bm{B}}_{\Theta_m}^{t+1}\bm{v}^j.
\end{align}
Then we convert \eqref{ProbV1} into the following sequential convex optimization form, where the subproblem at the $j$-th update is
\begin{subequations}
	\label{ProbV2}
	\begin{align}
	\label{ProbV21}
	\min_{\bm{v}}
	&\ \mu^t\bm{v}^{H}\hat{\bm{B}}_{\Theta_s}^{t+1}\bm{v}
	-2\mathrm{Re}\{\bm{v}^{H}\hat{\bm{B}}_{\Theta_m}^{t+1}\bm{v}^j\}
	+2\mathrm{Re}\{\bm{v}^{H}\hat{\bm{d}}^{t+1}\}   \\
	\label{ProbV22}
	{\rm{s.t.}} &\  \bm{v}^H \bm{\Pi}_n^{t+1} \bm{v} \le P_\mathrm{A}, \ \forall n, \\
	\label{ProbV23}
	&\ |v_l| \leq \varsigma, \ \forall l.
	\end{align}
\end{subequations}
As \eqref{ProbV2} is a convex problem, we can solve it directly by using off-the-shelf toolboxes, such as the interior-point (IP) method in CVX \cite{CVX2020}, or other solvers, such as ADMM \cite{Li2024Antenna}.

We summarize the CCCP-based algorithm to tackle the problem \eqref{ProbV1} in Algorithm 3, where $\iota_2$ is the termination tolerance and $J_{\max}$ is the maximum iteration number.

\subsection{Optimize $\mu$ with fixed $\bm{x}$ and $\bm{V}$}
According to its definition, the specific expression of $\mu$ at the $t$-th iteration is
\begin{align}
	\label{solvemu}
	\mu^{t+1} =\frac{(\bm{x}^{t+1})^H\bm{\Omega}_m^{t+1}\bm{x}^{t+1}} {(\bm{x}^{t+1})^H\bm{\Omega}_s^{t+1}\bm{x}^{t+1}}.
\end{align}
In reality, $\mu$ displays the convergent behavior of $\varGamma(\bm{x},\bm{V})$ after Dinkelbach transform.

We summarize the proposed beampattern synthesis algorithm for ARIS-aided MIMO radar in Algorithm 4, where $\iota_3$ and $T_{\max}$ are the termination tolerance and maximum iteration number.
By alternatingly optimizing the variables $\bm{x}$, $\bm{V}$ and $\mu$, the problem \eqref{Prob2} will converge to a locally optimal solution.

\begin{figure}[!t]
	\label{alg:AOxV}
	\begin{algorithm}[H]
		\caption{Proposed Beampattern Synthesis Algorithm.}
		\begin{algorithmic}[1]
			\renewcommand{\algorithmicrequire}{\textbf{Input:}}
			\REQUIRE $\Theta_m$, $\Theta_s$, $\bm{G}$, $P_A$, $\sigma_v^2$, $\iota_3$ and $T_{\max}$. \\         
			\renewcommand{\algorithmicrequire}{\textbf{Initialization:}}
			\REQUIRE $t=0$, Set $\bm{x}^{0}$, $\bm{V}^{0}$, $\mu^0=\frac{1}{\varGamma(\bm{x}^0,\bm{V}^0)}$.  
			\WHILE {$t \leq T_{\max}$ and $\vert \mu^{t+1}-\mu^{t} \vert>\iota_3$}
			\STATE Obtain $\bm{x}^{t+1}$ using Algorithm 2;
			\STATE Obtain $\bm{V}^{t+1}$ using Algorithm 3;
			\STATE Obtain $\mu^{t+1}$ using \eqref{solvemu};
			\STATE $t=t+1$;
			\ENDWHILE
			\renewcommand{\algorithmicensure}{\textbf{Output:}}
			\ENSURE MIMO transmit waveform $\bm{x}^{t+1}$ and ARIS reflection coefficients $\bm{V}^{t+1}$.  
		\end{algorithmic}
	\end{algorithm}
	\vspace{-10mm}
\end{figure}

\subsection{Convergence Analysis}
%

This subsection discusses the convergence of the proposed algorithm.
Note that the objective function $\varGamma(\bm{x},\bm{V})$ in \eqref{Prob1} is lower bounded since the matrices $\bm\Omega_s$ and $\bm\Omega_m$ are both positive definite.
We only need to prove that $\varGamma(\bm{x},\bm{V})$ is monotonically non-increasing after the updates of Algorithm 2, Algorithm 3 and Eq. \eqref{solvemu}, respectively \cite{Li2024Shaped}.



First, we prove that Algorithm 2 converges to a local optimal solution under some mild conditions as follows.

\textit{Theorem 1:} Let us assume $\lim_{k\to\infty}\Vert\bm{p}^{k+1}-\bm{p}^k\Vert_2=0$ and $\lim_{k\to\infty}\Vert\bm{r}^{k+1}-\bm{r}^k\Vert_2=0$.
With $\beta_1>0$, we define the sequence $\{\bm{x}^k,\bar{\bm x}^k,\hat{\bm x}^k,\bm{p}^k,\bm{r}^k\}$ is generated by Algorithm 2 for solving the subproblem \eqref{Probx1}.
Thus, there exists a stationary point $\{\bm{x}^\star,\bar{\bm x}^\star,\hat{\bm x}^\star,\bm{p}^\star,\bm{r}^\star\}$, which is the local optimal solution of~\eqref{Probx1}.

\textit{Proof:} It is easily ﻿derived similar to Theorem 1 in \cite{Ran2024Phase}.

Based on Theorem 1, we can obtain that $\varUpsilon(\bm{x}^{t+1},\bm{V}^{t},\mu^{t}) \leq \varUpsilon(\bm{x}^{t},\bm{V}^{t},\mu^{t})$.
Next, it is obvious that the objective function in \eqref{ProbV2} is monotonically non-increasing since \eqref{ProbV2} is a convex problem.
As a consequence, Algorithm 3 converges to a local optimal solution according to the principle of CCCP \cite{Sriperumbudur2009CCCP}. 
Then we obtain $\varUpsilon(\bm{x}^{t+1},\bm{V}^{t+1},\mu^{t}) \leq \varUpsilon(\bm{x}^{t+1},\bm{V}^{t},\mu^{t})$.

Based on the above analysis, we have $\varUpsilon(\bm{x}^{t+1},\bm{V}^{t+1},\mu^{t}) \leq \varUpsilon(\bm{x}^{t},\bm{V}^{t},\mu^{t})=0$.
Combining this inequality with the equation \eqref{solvemu}, we obtain $\mu^t \leq \mu^{t+1}$. 
Since $\mu=\frac{1}{\varGamma(\bm{x},\bm{V})}$, we have $\varGamma(\bm{x}^{t+1},\bm{V}^{t+1}) \leq \varGamma(\bm{x}^t,\bm{V}^t)$.
That is to say, $\varGamma(\bm{x},\bm{V})$ is monotonically non-increasing in Algorithm 4.  
Therefore, the proposed algorithm is convergent when the scaled-dual variables satisfy $\lim_{k\to\infty}\Vert\bm{p}^{k+1}-\bm{p}^k\Vert_2=0$ and $\lim_{k\to\infty}\Vert\bm{r}^{k+1}-\bm{r}^k\Vert_2=0$.


\subsection{ Computational Complexity}

This subsection analyses the computational complexity of the proposed algorithm.
The complexity of Algorithm 4 mainly depends on the Algorithm 2, Algorithm 3 and \eqref{solvemu}.
Obviously, the complexity of Algorithm 2 hinges on the solutions to the subproblems \eqref{ADMM:sub1a}, \eqref{ADMM:sub2a} and \eqref{ADMM:sub3a}.
Specifically, the cost of solving \eqref{ADMM:sub1a} is determined primarily by the inverse of the matrix$({\bm I}_{L_1} +\varepsilon^{\star}\bm{R})$ and equals $\mathcal{O}(L_1^{3})$.
The costs of solving \eqref{ADMM:sub2a} for all three kinds of waveform constraints are the same and equal $\mathcal{O}{(N L_1)}$.
The cost of solving \eqref{ADMM:sub2a} is dominated by the inverse of the matrix $(\beta_1\bm{I}_{NL_1}+\mu^t\bm{\Omega}^t_s-\bm{\Omega}^t_m)$ and equals $\mathcal{O}(N^{3}L_1^{3})$.
Therefore, the total cost of Algorithm 2 is $\mathcal{O}(K_{x}N^{3}L_1^{3})$, where $K_{x}$ is the iteration number of Algorithm 2.
Note that the cost of IP method solving \eqref{ProbV2} is $\mathcal{O}((N^3 L_2^{0.5}+N^2 L_2^{2.5}+N L_2^{3.5}+L_2^{4.5})\log_{}{(1/{\varepsilon})})$, where $\varepsilon$ is the accuracy of solution.
Then the total cost of Algorithm 3 equals $\mathcal{O}(J_{v} (N^3 L_2^{0.5}+N^2 L_2^{2.5}+N L_2^{3.5}+L_2^{4.5})\log_{}{(1/{\varepsilon})})$, where $J_{v}$ is the iteration number of Algorithm 3.
Obviously, the calculation cost of \eqref{solvemu} equals $\mathcal{O}(N^{2}L_1^{2})$.
In summary, the complexity of Algorithm 4 in each iteration is $\mathcal{O}(K_{x} N^{3} L_1^{3}+J_{v}(N^3 L_2^{0.5}+N^2 L_2^{2.5}+N L_2^{3.5}+L_2^{4.5})\log_{}{(1/{\varepsilon})})$.

\section{Numerical Results and Analysis}
This section examines the beampattern synthesis performance of ARIS-aided MIMO radar via several numerical experiments.
The beampattern of conventional MIMO radar (RIS-free) is also given as a benchmark. 
In fact, without the assistance of ARIS, the proposed problem degenerates into the problem in \cite{Cheng2018Communication} after removing the spectral and similarity constraints.
Even though the algorithm in \cite{Cheng2018Communication} adopts a different solving scheme from the proposed algorithm, both two algorithms always have identical beampattern performance.
We assume that the MIMO transmit array is located at the coordinate origin, the ARIS is located at $(-1.94\mathrm{m},0.5\mathrm{m})$, the angle and distance between ARIS and MIMO array are respectively equal to $\theta_p=10^{\circ}$.
The distance-dependent path loss is modelled as $PL=PL_0(\frac{D}{D_0})^{-\alpha}$, where $PL_0=-30\text{dB}$ is the path loss at the reference distance $D_0 = 1\mathrm{m}$, 
$D=2\mathrm{m}$ is the distance between ARIS and MIMO array, and $\alpha = 2.2$ is the path loss exponent. 
The channel $\bm{G}$ follows a Rician fading model \cite{Wu2019Intelligent}.
In the experiments, we set the Rician factor $K_\mathrm{R}=3$, the incident noise variance $\sigma_v^2=-80 \ {\text{dBm}}$ and the maximum available power of ARIS $P_\mathrm{A} = 1 \mathrm{W}$.
The inter-element spaces of MIMO array and ARIS are both half-wavelength.
Unless otherwise specified, we set $L_1=10$, $N=32$, $L_2 = 64$ and $\varsigma = 5$, respectively.
The subproblem (49) is solved by the IP method via CVX.
Besides, all numerical experiments perform on a standard PC with an Intel  Core i7 3.1 GHz CPU and 64 GB RAM.



\subsection {Single-mainlobe beampattern}


\begin{figure}[!t]
	\centering
	\includegraphics[width=0.5\linewidth]{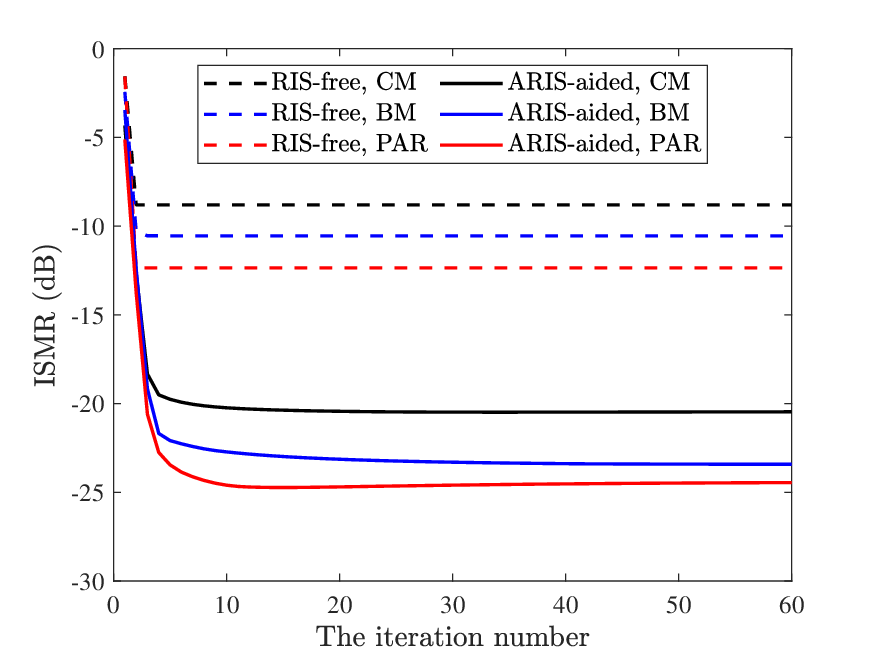}
	\caption{Convergence behaviors of single-mainlobe beampattern synthesis under CM, BM and PAR waveform constraints.}
	\label{Fig:ISMR_single}   
	\vspace{-6mm}
\end{figure}

\begin{figure}[!t]
	\centering
	\includegraphics[width=0.5\linewidth]{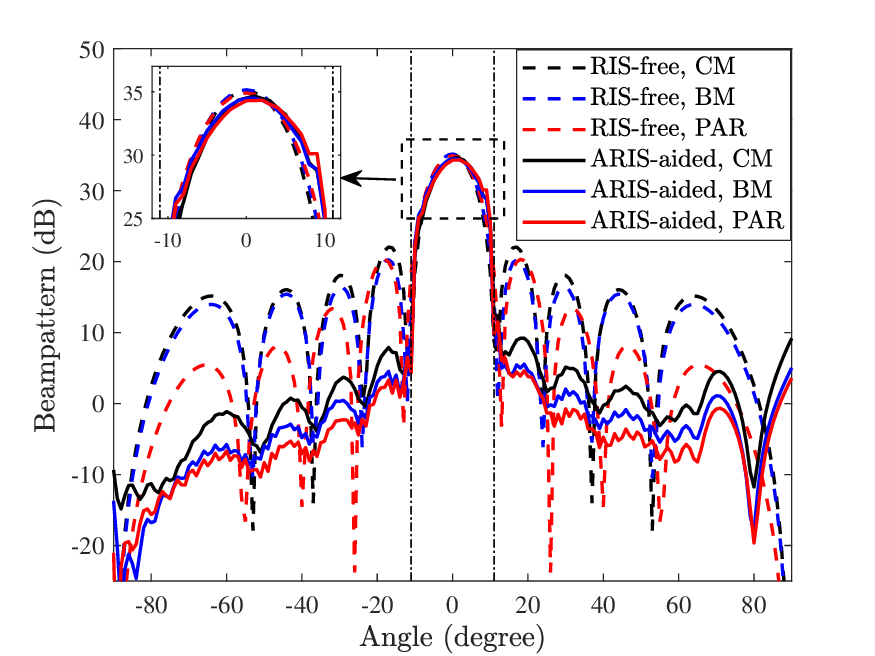}
	\caption{Single-mainlobe beampatterns under CM, BM and PAR waveform constraints.}
	\label{Fig:beampattern_single}   
	\vspace{-2mm}
\end{figure}

This subsection investigates the single-mainlobe beampattern performance of ARIS-aided MIMO radar under different waveform constraints. 
For the PAR and BM constraints, we set $\eta=1.2$ and $\delta=0.1$, respectively.
The mainlobe and sidelobe regions are $\Theta_m=[-11^{\circ},11^{\circ}]$ and $\Theta_s=[-90^{\circ},-11^{\circ}]\cup[11^{\circ},90^{\circ}]$.
Fig. \ref{Fig:ISMR_single} illustrates the variation curves of ISMR with respect to the iteration number of the proposed algorithms for the ARIS-aided and conventional MIMO radars, respectively.
We observe that all ISMRs rapidly converge to a stationary value after a few number of iterations for ARIS-aided MIMO radars, while the iterative number of conventional ones equal only one since the AM is not required.
The deployment of ARIS dramatically reduces ISMRs for different waveform constraints.
Compared to conventional MIMO radars, the ARIS-aided ones reduce the ISMR to $-20.45 \mathrm{dB}$, $-23.42 \mathrm{dB}$ and $-24.40 \mathrm{dB}$, respectively, under CM, BM and PAR constraints.
The looser the waveform constraint, the better the beampattern performance.
The ISMRs are improved by $11.64 \mathrm{dB}$, $12.87 \mathrm{dB}$ and $12.04 \mathrm{dB}$, respectively.

\begin{table*}[!t]
	\centering
	\caption{The energy distribution of single-mainlobe beampatterns}
	\begin{tabular}{|c|cc|cc|cc|}
		\hline
		\multirow{2}{*}{Waveform} & \multicolumn{2}{c|}{Sidelobe ($\mathrm{dB}$)}    & \multicolumn{2}{c|}{Mainlobe ($\mathrm{dB}$)}     & \multicolumn{2}{c|}{ISMR ($\mathrm{dB}$)}            \\ \cline{2-7} 
		& \multicolumn{1}{c|}{RIS-free} & ARIS-aided & \multicolumn{1}{c|}{RIS-free} & ARIS-aided & \multicolumn{1}{c|}{RIS-free} & ARIS-aided \\ \hline
		CM                & \multicolumn{1}{c|}{36.42}    & 24.80      & \multicolumn{1}{c|}{45.23}    & 45.25      & \multicolumn{1}{c|}{-8.81}    & -20.45     \\ \hline
		BM                & \multicolumn{1}{c|}{34.96}    & 21.89      & \multicolumn{1}{c|}{45.51}    & 45.31      & \multicolumn{1}{c|}{-10.55}   & -23.42     \\ \hline
		PAR               & \multicolumn{1}{c|}{33.09}    & 20.84      & \multicolumn{1}{c|}{45.45}    & 45.24      & \multicolumn{1}{c|}{-12.36}   & -24.40     \\ \hline
	\end{tabular}
	\vspace{-6mm}
\end{table*}


\begin{figure}[!t]
	\centering
	\includegraphics[width=0.5\linewidth]{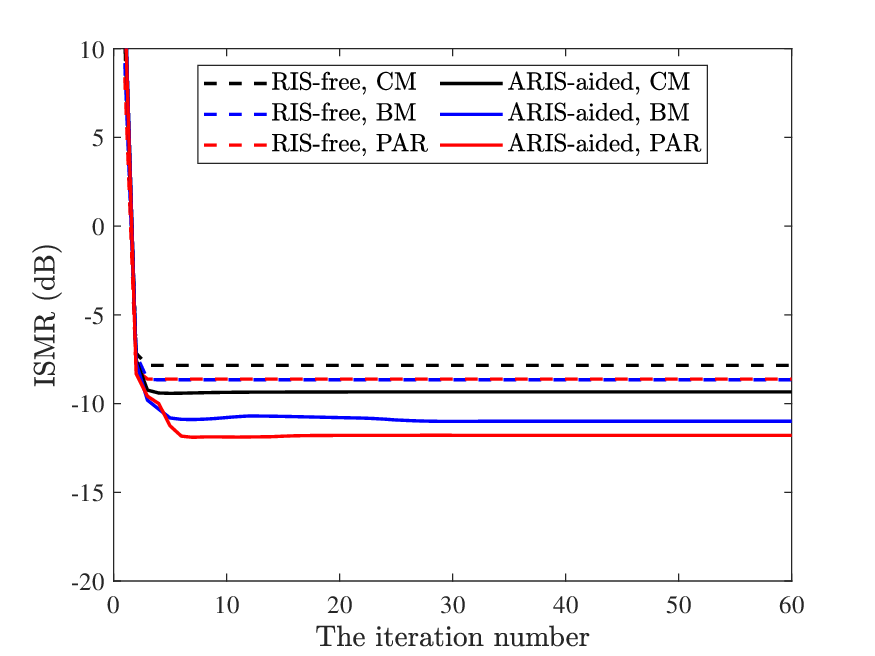}
	\caption{Convergence behaviors of double-mainlobe beampattern synthesis under CM, BM and PAR waveform constraints.}
	\label{Fig:ISMR_MUL}   
\end{figure}

\begin{table*}[!tbp]
	\centering
	\caption{The consumed CPU time of single-mainlobe beampatterns}	
	\vspace{-4mm}
	\begin{tabular}{ccccc}
		\multicolumn{1}{l}{}                            & \multicolumn{1}{l}{}                           & \multicolumn{1}{l}{}          & \multicolumn{1}{l}{}          & \multicolumn{1}{l}{}            \\ \hline
		\multicolumn{1}{|c|}{\multirow{3}{*}{Waveform}} & \multicolumn{4}{c|}{CPU time (s)}                                                                                                                \\ \cline{2-5} 
		\multicolumn{1}{|c|}{}                          & \multicolumn{1}{c|}{\multirow{1}{*}{RIS-free}} & \multicolumn{3}{c|}{ARIS-aided}                                                                 \\ \cline{3-5} 
		\multicolumn{1}{|c|}{}                          & \multicolumn{1}{c|}{(Proposed/[16])}                          & \multicolumn{1}{c|}{update $\bm{x}$} & \multicolumn{1}{c|}{update $\bm{V}$} & \multicolumn{1}{c|}{total time} \\ \hline
		\multicolumn{1}{|c|}{CM}                        & \multicolumn{1}{c|}{7.45/106.57}                      & \multicolumn{1}{c|}{13.96}    & \multicolumn{1}{c|}{209.63}   & \multicolumn{1}{c|}{223.59}     \\ \hline
		\multicolumn{1}{|c|}{BM}                        & \multicolumn{1}{c|}{6.42/$\times$}                      & \multicolumn{1}{c|}{16.59}    & \multicolumn{1}{c|}{261.62}   & \multicolumn{1}{c|}{278.21}     \\ \hline
		\multicolumn{1}{|c|}{PAR}                       & \multicolumn{1}{c|}{16.17/106.58}                     & \multicolumn{1}{c|}{42.56}    & \multicolumn{1}{c|}{189.93}   & \multicolumn{1}{c|}{232.49}     \\ \hline
	\end{tabular}
\vspace{-6mm}
\end{table*}


Fig. \ref{Fig:beampattern_single} presents the single-mainlobe beampatterns of ARIS-aided and conventional MIMO radars corresponding to Fig. 2.
Evidently, the ARIS significantly reduces the energy in sidelobe region, regardless of the type of waveform constraints.
In contrast, it hardly changes the energy in mainlobe region, but only makes the energy distribution more even.
That is primarily because the ARIS reflection signal has only limited energy.
From the perspective of improving ISMR performance, it is more efficient to reduce the sidelobe energy than to enhance the mainlobe energy.
Table I shows the specific energy changes in the sidelobe and mainlobe regions.
For the CM, BM and PAR constraints, the sidelobe energies decrease by $11.62 \rm{dB}$, $13.07 \rm{dB}$ and $12.25 \rm{dB}$, respectively, while the mainlobe energies has almost no increase but a slight decrease in the latter two constraints.
The closely-equipped ARIS has a powerful ability to boost beampattern performance of MIMO radars by remarkably weakening the sidelobe energy, and it is suitable for various types of practical waveforms.

Table II displays the consumed CPU times of the proposed algorithm in Fig. 3.
We also provide the CPU times of RIS-free MIMO radars which are realized by the simplified variants of the proposed algorithm and the algorithm in \cite{Cheng2018Communication}, respectively.
Note that the algorithm in \cite{Cheng2018Communication} is not applicable to handle the BM constraint. 
We only obtain its CPU times under CM or PAR constraints.
For the proposed algorithm, ARIS-aided radars spend significantly more CPU times than RIS-free ones, regardless of the waveform constraint.
By comparing the CPU times of updating $\bm{x}$ and $\bm{V}$, we find that the update of $\bm{V}$ dominates the total computational overhead. 
This is because the subproblem (49) is solved by the IP method for the update of $\bm{V}$ in (47), which requires relatively large computational cost.
Therefore, a feasible avenue to speed up the proposed algorithm is to replace the IP method with other fast solvers.
Compared with the algorithm in \cite{Cheng2018Communication} for RIS-free radars, the proposed algorithm has overwhelmingly less CPU time, and it expends about only twice the CPU time even if working for ARIS-aided radars.


\subsection{Double-mainlobe beampattern}


\begin{figure}[!t]
	\centering
	\includegraphics[width=0.5\linewidth]{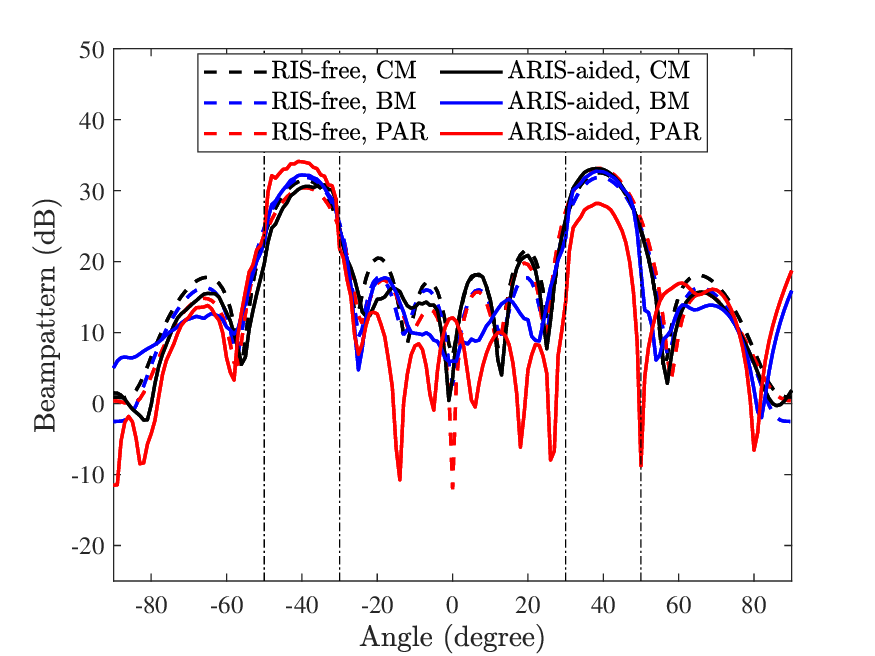}
	\caption{Double-mainlobe beampatterns under CM, BM and PAR waveform constraints.}
	\label{Fig:beampattern_MUL}   
\end{figure}

This subsection considers the double-mainlobe beampattern performance under different waveform constraints. 
The waveform parameters are the same as those in Subsection IV.A.
The mainlobe and sidelobe regions are respectively set as $\Theta_m=[-51^{\circ},-29^{\circ}]\cup[29^{\circ},51^{\circ}]$ and $\Theta_s=[-90^{\circ},51^{\circ}]\cup[-29^{\circ},29^{\circ}]\cup[51^{\circ},90^{\circ}]$.
Fig. \ref{Fig:ISMR_MUL} demonstrates the convergence curves of the proposed algorithm for the ARIS-aided and conventional MIMO radars. 
Similar to Fig. \ref{Fig:ISMR_single}, the proposed algorithm rapidly converges to a stationary value.
Under the same waveform constraint, the ISMR of ARIS-aided MIMO radar is always less than that of conventional one.
But the improvement in ISMR decreases visibly.
Concretely, deploying an ARIS reduces ISMRs to $-10.59 \mathrm{dB}$, $-10.99 \mathrm{dB}$ and $-11.79 \mathrm{dB}$ for the CM, BM and PAR constraints, respectively, and the ISMRs are only improved by $2.74 \mathrm{dB}$, $2.33 \mathrm{dB}$ and~$3.17 \mathrm{dB}$.

\begin{table*}[!tbp]
	\centering
	\caption{The energy distribution of double-mainlobe beampatterns}
	\begin{tabular}{|c|cc|cc|cc|}
		\hline
		\multirow{2}{*}{Waveform} & \multicolumn{2}{c|}{Sidelobe ($\mathrm{dB}$)}     & \multicolumn{2}{c|}{Mainlobe ($\mathrm{dB}$)}     & \multicolumn{2}{c|}{ISMR ($\mathrm{dB}$)}   \\ \cline{2-7} 
		& \multicolumn{1}{c|}{RIS-free} & ARIS-aided & \multicolumn{1}{c|}{RIS-free} & ARIS-aided & \multicolumn{1}{c|}{RIS-free} & ARIS-aided \\ \hline
		CM                & \multicolumn{1}{c|}{38.40}    & 36.04      & \multicolumn{1}{c|}{46.25}    & 46.63     & \multicolumn{1}{c|}{-7.85}    & -10.59     \\ \hline
		BM                & \multicolumn{1}{c|}{37.56}    & 35.72      & \multicolumn{1}{c|}{46.22}    & 46.71      & \multicolumn{1}{c|}{-8.66}   & -10.99     \\ \hline
		PAR               & \multicolumn{1}{c|}{37.75}    & 34.55      & \multicolumn{1}{c|}{46.37}    & 46.34      & \multicolumn{1}{c|}{-8.62}   & -11.79     \\ \hline
	\end{tabular}
	\vspace{-6mm}
\end{table*}

\begin{table*}[!t]
	\centering
	\caption{The consumed CPU time of double-mainlobe beampatterns}	
	\vspace{-4mm}
	\begin{tabular}{ccccc}
		\multicolumn{1}{l}{}                            & \multicolumn{1}{l}{}                           & \multicolumn{1}{l}{}          & \multicolumn{1}{l}{}          & \multicolumn{1}{l}{}            \\ \hline
		\multicolumn{1}{|c|}{\multirow{3}{*}{Waveform}} & \multicolumn{4}{c|}{CPU time (s)}   
		\\ \cline{2-5} 
		\multicolumn{1}{|c|}{}                          & \multicolumn{1}{c|}{\multirow{1}{*}{RIS-free}} & \multicolumn{3}{c|}{ARIS-aided}  
		\\ \cline{3-5} 
		\multicolumn{1}{|c|}{}                          & \multicolumn{1}{c|}{(Proposed/[16])}                          & \multicolumn{1}{c|}{update $\bm{x}$} & \multicolumn{1}{c|}{update $\bm{V}$} & \multicolumn{1}{c|}{total time} \\ \hline
		\multicolumn{1}{|c|}{CM}                        & \multicolumn{1}{c|}{5.07/111.37}                      & \multicolumn{1}{c|}{22.73}    & \multicolumn{1}{c|}{202.51}   & \multicolumn{1}{c|}{225.25}     \\ \hline
		\multicolumn{1}{|c|}{BM}                        & \multicolumn{1}{c|}{5.68/$\times$}                      & \multicolumn{1}{c|}{23.16}    & \multicolumn{1}{c|}{218.39}   & \multicolumn{1}{c|}{241.56}     \\ \hline
		\multicolumn{1}{|c|}{PAR}                       & \multicolumn{1}{c|}{43.81/112.03}                     & \multicolumn{1}{c|}{40.40}    & \multicolumn{1}{c|}{235.49}   & \multicolumn{1}{c|}{275.89}     \\ \hline
	\end{tabular}
	\vspace{-6mm}
\end{table*}


Fig. \ref{Fig:beampattern_MUL} shows the double-mainlobe beampatterns of ARIS-aided and conventional MIMO radars corresponding to Fig. \ref{Fig:ISMR_MUL}.
It is observed that the assistance of ARIS still effectively reduces SLLs.
But the decrease in SLLs is remarkably smaller than that in the single-mainlobe case.
Table III gives the specific energy changes in the sidelobe and mainlobe regions.
With the help of ARIS, the sidelobe energies are respectively reduced by $2.36 \mathrm{dB}$, $1.84 \mathrm{dB}$ and $3.20 \mathrm{dB}$ for the CM, BM and PAR constraints, while the mainlobe energies increase by $0.38 \mathrm{dB}$, $0.49 \mathrm{dB}$ and $-0.03 \mathrm{dB}$.
There seems to be a trade-off between the energies of mainlobe and sidelobe regions.
Note that there exists an energy imbalance phenomenon within the two main lobes. 
The looser the waveform constraints, the more severe the imbalance in two main lobes.
The reason is chiefly that the metric in \eqref{ISMR} does not impose constraints on the energy allocation in the two main lobes and the greater waveform DoF will exacerbate this phenomenon.
To remedy this issue, a feasible method is to replace the denominator of ISMR in \eqref{ISMR} with the energy of each main lobe.
Then we can adopt the maximum of two modified ISMRs as the metric and customize a specialized algorithm to tackle this min-max FP problem.

Table IV presents the consumed CPU times of all beampatterns in Fig. 5.
Similar to Table II, the proposed algorithm spends strikingly more CPU time for ARIS-aided radars than for RIS-free radars, and it has extremely superior performance on CPU time compared to the algorithm in \cite{Cheng2018Communication}.
Even though the number of mainlobes increases to 2, the consumed CPU times are basically comparable to those in Table II since the ISMR is robust to the setting of mainlobe region.


\begin{figure}[t!]
	\vspace{-3mm}
	\centering
	\subfloat[]{\includegraphics[width=0.5\linewidth]{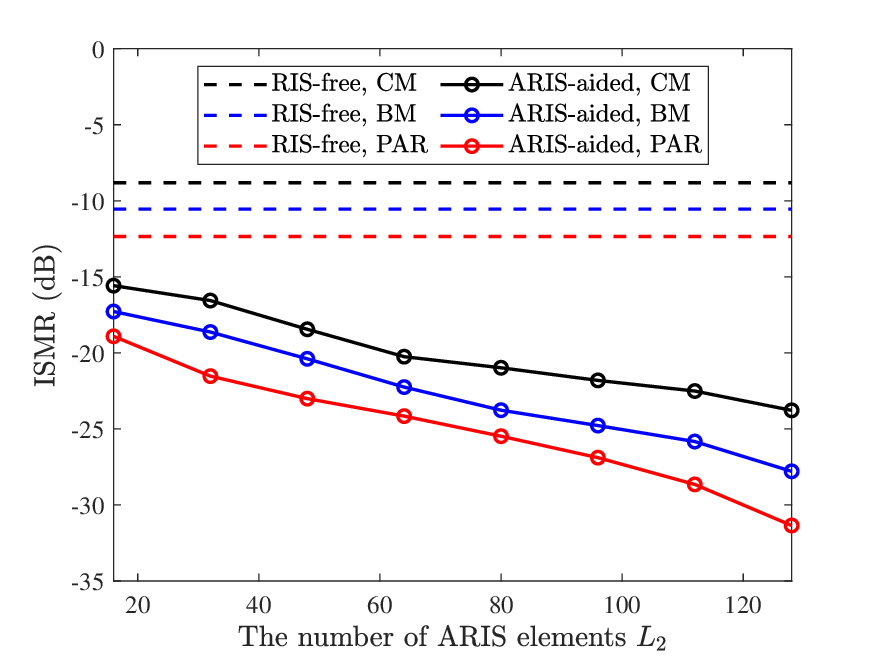}		\label{Fig:L2_16_128}} 
	\subfloat[]{\includegraphics[width=0.5\linewidth]{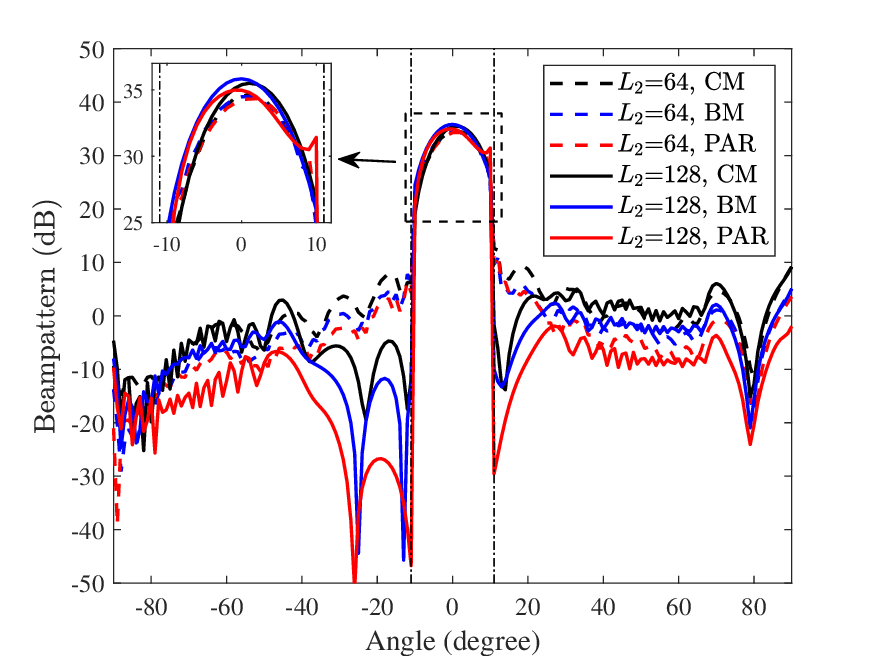}
	\label{Fig:L2_64_128_beampattern}}
	\caption{ISMR versus the element number of ARIS. (a) ISMR, (b) the beampatterns with $L_2=64$ and $L_2=128$.}
	\label{Fig:L2}
	\vspace{-8mm}
\end{figure}

\begin{figure}[t!]
	\vspace{-3mm}
	\centering
	\subfloat[]{\includegraphics[width=0.5\linewidth]{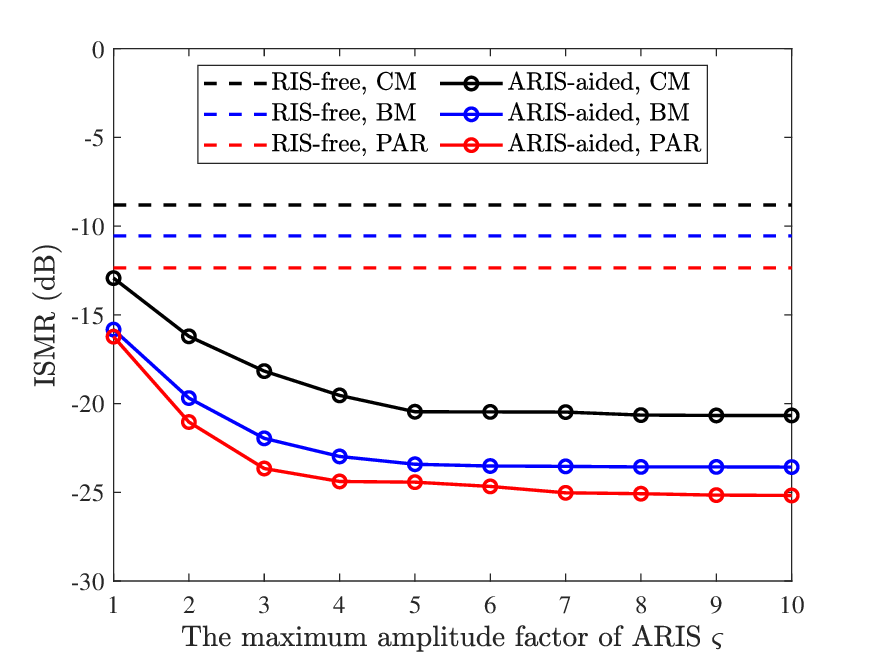}
	\label{Fig:varsigma_1_10}}
	\subfloat[]{\includegraphics[width=0.5\linewidth]{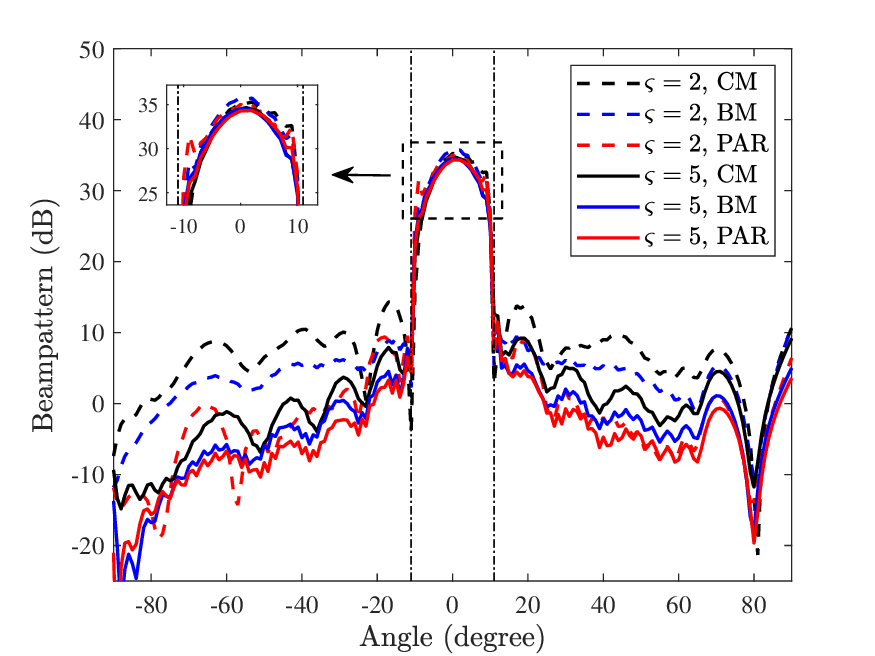}
	\label{Fig:varsigma_5_10_beampattern}}
	\caption{ISMR versus the maximum amplification factor of ARIS. (a) ISMR, (b) the beampatterns with $\varsigma=2$ and $\varsigma=5$.}
    \label{Fig:varsigma}
	\vspace{-6mm}
\end{figure}



\subsection{The impact of ARIS parameters}

This subsection discusses the influence of two ARIS parameters, the element number $L_2$ and the maximum amplification factor $\varsigma$, on the single-mainlobe beampattern performance.
In the following experiments, the mainlobe center $\theta_0$ is randomly uniformly distributed in $[-10^{\circ},10^{\circ}]$. 
The mainlobe and sidelobe regions are respectively set as $\Theta_m=[\theta_0-11^{\circ},\theta_0+11^{\circ}]$ $\Theta_s=[-90^{\circ},\theta_0-11^{\circ}]\cup[\theta_0+11^{\circ},90^{\circ}]$. 
All the ISMRs are calculated by averaging 20 independent runs.

Fig. \ref{Fig:L2}\subref{Fig:L2_16_128} displays the variation curves of ISMRs with respect to the ARIS element number under different waveform constraints.
We observe that the ARIS-aided MIMO radars always have significantly lower ISMRs than the conventional ones for all three waveforms, and the ISMRs markedly decrease with the increase of $L_2$.
Comparing the three waveform constraints, the PAR constraint provides the best ISMR performance.
The performance gaps among the three constraints slightly widen as the number of ARIS elements increases.
Fig. \ref{Fig:L2}\subref{Fig:L2_64_128_beampattern} shows the beampatterns under three different waveform constraints with $L_2=64$ and $L_2=128$, where $\theta_0=0^\circ$. 
We find that the greater number of elements leads to the deeper reduction of SLLs, especially in the areas adjacent to the mainlobe region, hence resulting in the better ISMR performance.

Fig. \ref{Fig:varsigma}\subref{Fig:varsigma_1_10} shows the variation curves of ISMRs as a function of the maximum amplification factor $\varsigma$.
We observe that all the ISMRs gradually decrease with the increase of $\varsigma$ when $\varsigma \leq 5$ and subsequently reach saturation when $\varsigma > 5$.
Concretely, the ISMRs respectively decrease by $7.53 \rm{dB}$, $7.57 \rm{dB}$, and $8.20 \rm{dB}$ for the CM, BM and PAR constraints when $\varsigma$ increases from 1 to 5, while they only decrease by $0.20 \rm{dB}$, $0.06 \rm{dB}$, and $0.51 \rm{dB}$ when $\varsigma$ increases from 6 to 10.
This is because the ARIS reflection coefficients $\bm{v}$ are limited by \eqref{RISpower} and \eqref{RISamplification} simultaneously and thus will be dominated by the maximum reflecting power of ARIS in \eqref{RISpower} if $\varsigma$ is large enough.
In this case, we have to increase the maximum reflecting power to ensure the effectiveness of $\varsigma$. 
Fig. \ref{Fig:varsigma}\subref{Fig:varsigma_5_10_beampattern} displays the corresponding beampatterns with $\varsigma=2$ and $\varsigma=5$, where $\theta_0=0^\circ$. 
Similar to Fig. \ref{Fig:beampattern_single}, as the maximum amplification factor of ARIS increases, the SLLs significantly decrease and the mainlobe energy has only a slight change.
In practice, we can select the ARIS parameters appropriately according to the requirement of beampattern performance.

\subsection{The impact of waveform parameters}

\begin{figure}[t!]
	\vspace{-3mm}
	\centering
	\subfloat[]{\includegraphics[width=0.5\linewidth]{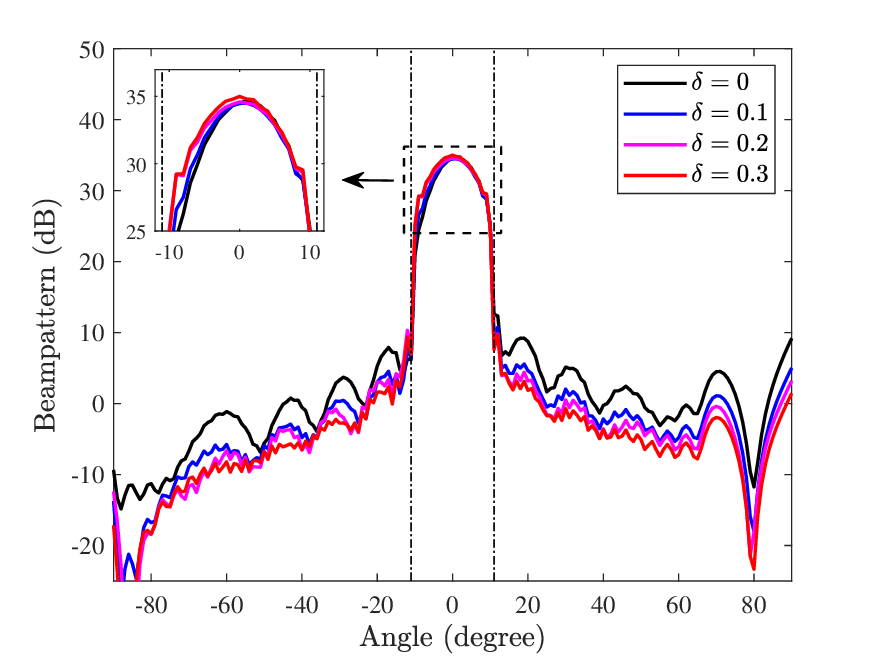}
		\label{Fig:delta}}
	\subfloat[]{\includegraphics[width=0.5\linewidth]{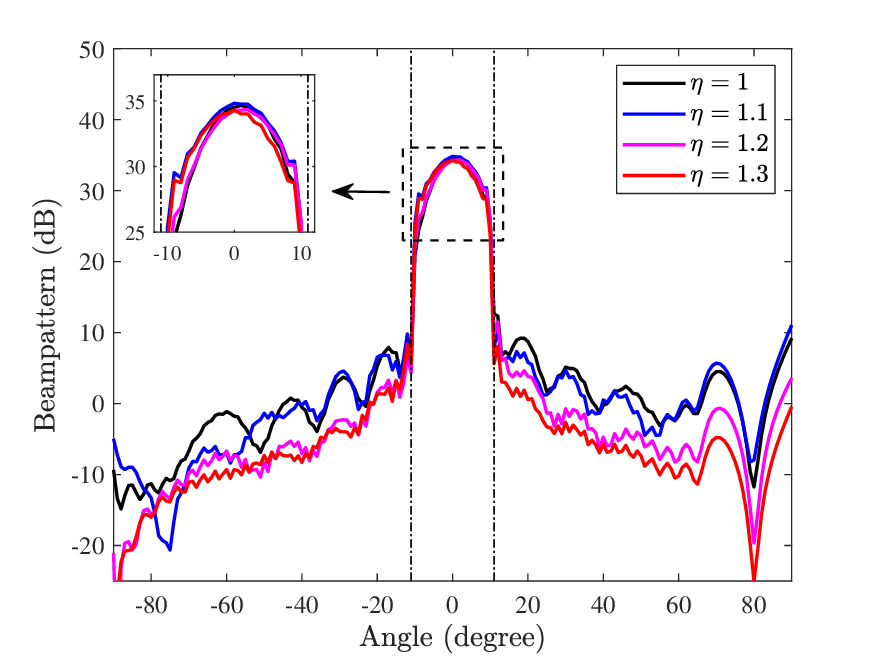}
		\label{Fig:epsilon}}
	\caption{The beampatterns for different BM and PAR parameters. (a) $\delta$ in BM constraints, (b) $\eta$ in PAR constraints.}
    \label{Fig:parameters}
	\vspace{-6mm}
\end{figure}

This subsection tests the influence of waveform parameters in BM and PAR constraints on the single-mainlobe beampattern performance.
Fig. \ref{Fig:parameters}\subref{Fig:delta} presents the BM-constrained beams with different parameters $\delta$. 
When $\delta=0$, the BM constraint is equivalent to the CM constraint.
As the value of $\delta$ increases, i.e., the dynamic range of waveforms grows up, all the SLLs decrease remarkably and the mainlobe energy increases mildly. 
But the rate of decrease in the SLLs gradually diminishes.
The corresponding ISMRs equal $-20.40 \mathrm{dB}$, $-23.42 \mathrm{dB}$, $-24.51 \mathrm{dB}$ and $-25.71 \mathrm{dB}$ for $\delta=0$, $0.1$, $0.2$ and $0.3$, respectively.
Fig. \ref{Fig:parameters}\subref{Fig:epsilon} displays the PAR-constrained beams with different parameters $\eta$.
The PAR constraint also degenerates into the CM constraint when $\eta =1$.
Similar to Fig. \ref{Fig:parameters}\subref{Fig:delta}, a larger value of $\eta$ implies a greater DoF of transmit waveform.
In Fig. \ref{Fig:parameters}\subref{Fig:epsilon}, it is observed that the larger the value of $\eta$, the lower the SLLs and the sightly higher the mainlobe energy. 
Only a marginal reduction on the SLL appears when $\eta \geq 1.2$, leading to a tiny improvement of ISMR.
Therefore, we have to properly choose the parameters of BM or PAR constrained waveform to make a good balance between beampattern performance and transmit efficiency.

\section{Conclusion}
This paper discussed the beampattern synthesis of ARIS-aided colocated MIMO radar.
We focused on maximizing the ISMR under the constraints of practical waveforms and ARIS's reflecting power and maximum amplification factor, and coined it as a nonconvex constrained FP problem. 
To solve this problem effectively, we first equivalently expressed the fractional objective function as an integral form by leveraging Dinkelbach transform, and then designed the practical waveforms and ARIS reflection coefficients under the framework of alternating minimization.
We optimized the CM, PAR or BM waveforms by customizing a CADMM-based algorithm in which the optimal solutions of all subproblems are obtained, and updated the ARIS reflection coefficients by capitalizing on a CCCP-based algorithm which only contains several simple convex subproblems.
Owing to additional DoFs offered by the ARIS, the proposed ARIS-aided MIMO radar can significantly reduce the energy in sidelobe region with an essentially unchanged energy in mainlobe region, resulting in superior ISMR performance than conventional MIMO radars.
Its ISMR performance is further improved with the increasing of ARIS element number and maximum amplification factor.



\bibliographystyle{IEEEtran}
\bibliography{ARIS_MIMO_ISMR}

\end{document}